\documentclass[conference]{IEEEtran}
\IEEEoverridecommandlockouts
\usepackage{cite}
\usepackage{amsmath,amssymb,amsfonts}
\usepackage{algorithm}
\usepackage{algpseudocode}
\usepackage{graphicx}
\usepackage{textcomp}
\usepackage[table]{xcolor}
\usepackage{hyperref}
\usepackage{comment}
\usepackage{soul}
\usepackage[caption=false,font=footnotesize]{subfig}
\def\BibTeX{{\rm B\kern-.05em{\sc i\kern-.025em b}\kern-.08em
    T\kern-.1667em\lower.7ex\hbox{E}\kern-.125emX}}

\usepackage{booktabs}
\usepackage{multirow}
\usepackage{adjustbox}
\begin{document}

\title{Exploiting the Interplay of Compute- and Memory-Bound kernels in MPI Applications}
\author{\IEEEauthorblockN{Ayesha Afzal}
\IEEEauthorblockA{Erlangen National High Performance\\ Computing Center (NHR@FAU)\\
Erlangen, Germany\\
ORCID: \href{https://orcid.org/0000-0001-5061-0438}{0000-0001-5061-0438}\\
Email: ayesha.afzal@fau.de}
\and
\IEEEauthorblockN{Krishna Manda}
\IEEEauthorblockA{University of Bonn\\ Germany\\
ORCID: \href{https://orcid.org/0000-0002-9744-3080}{0000-0002-9744-3080}\\
Email: kmanda@uni-bonn.de}
\and
\IEEEauthorblockN{Georg Hager}
\IEEEauthorblockA{Erlangen National High Performance\\ Computing Center (NHR@FAU)\\
Erlangen, Germany\\
ORCID: \href{https://orcid.org/0000-0002-8723-2781}{0000-0002-8723-2781}\\
Email: georg.hager@fau.de}}
\maketitle

\begin{abstract}
Parallel applications are often designed for synchronous, lock-step execution, treating communication stalls as performance hazards. Yet, in a communication-light application without frequent synchronization points that alternates between compute-bound  memory-bound execution, an MPI communication stall can act as an unintentional relief on memory-bandwidth contention. 
We demonstrate this using a Parallel Optical Flow Solver, which combines a compute-bound \emph{Ray Tracing} kernel with a memory-bound \emph{Optical Flow Solver} kernel and negligible inter-process communication. This program shows considerable speedup via desynchronization and automatic overlap between compute- and memory-bound phases, showing that natural desynchronization is an architecture-aware optimization.
An optimal speedup is achieved when the number of processes concurrently executing the memory-bound phase on a ccNUMA domain is near the bandwidth saturation point.
We also show a case where reducing communication overhead using MPI asynchronous progress significantly degrades performance because it allows too many ranks to contend for memory bandwidth simultaneously. 
In order to study the dynamics under more controlled conditions, we develop a tunable dual-kernel microbenchmark,
with which we show that significant application or system noise (natural or injected) is required to achieve full desynchronization. Finally, we also validate these results using a bandwidth-aware, model-based simulator. 

\end{abstract}


\section{Introduction and related work}
High-Performance Computing (HPC) applications are often designed with synchronous, lock-step execution, alternating between computation and communication phases. On real systems, this ideal is rarely achieved: Workload imbalance, system noise, and bottlenecks cause processes to diverge, a behavior known as \emph{desynchronization}~\cite{afzal2021analytic}. It is usually considered a performance hazard, since a stalled MPI process means idle hardware, and the standard engineering response is to remove whatever is causing the stall. This paper shows a case where this assumption can fail. In a communication-light multi-kernel application combining compute-bound and memory-bound kernels, an MPI communication stall can act as an unintentional relief on memory-bandwidth contention; removing it by any means can reduce throughput. This behavior arises because compute-bound and memory-bound kernels scale fundamentally differently: Compute-bound phases continue to benefit from additional cores, whereas memory-bound phases quickly saturate the shared bandwidth on a ccNUMA domain. Their interaction can therefore create spontaneous desynchronization even when communication is negligible. In addition, desynchronization can provide a performance boost by preventing excessive simultaneous demand on shared bottlenecks~\cite{afzal2022role}.

\paragraph*{Related Work} Prior research on desynchronization has focused mainly on communication-heavy workloads, where synchronization delays propagate as \emph{idle waves}~\cite{afzal2019propagation,losada2019asynchronous,markidis2015idle} and analytic models describe their evolution~\cite{afzal2021analytic}.
Studies of large-scale applications have also analyzed synchronization inefficiencies and wait states~\cite{bohme2016identifying}, while resilience techniques for extreme-scale stencil applications address execution disruptions through local recovery and failure masking~\cite{gamell2015local}. Asynchronous update methods further explore relaxing lock-step execution constraints to improve scalability~\cite{kolakowska2004desynchronization}
Related performance models include the Execution-Cache-Memory (ECM) model~\cite{treibig2009bandwidth}, the Roofline model~\cite{williams2009roofline}, and communication-cost models such as Hockney's model~\cite{hockney1994}. These models support the analysis on real hardware or simulation via DisCostiC~\cite{AfzalPoster:2024,AfzalPoster:2025}, but they primarily target single-kernel or communication-driven behavior. In memory-bound workloads, prior work introduced the concept of \emph{saturation cores}, showing that once too many cores compete for shared memory bandwidth, performance degrades regardless of
available compute capacity, and that desynchronization can emerge naturally from bandwidth variation across ccNUMA domains~\cite{afzal2020desync}. Other studies investigate communication-computation overlap using non-blocking communication~\cite{AfzalHW:2023:1,afzal2022exploring}, but communication remains the primary source of desynchronization.

\paragraph*{Research Gap} This phenomenon extends beyond single-kernel solvers. Many HPC and simulation pipelines combine compute-intensive and memory-intensive phases, such as rendering integrated with physics simulations or machine learning preprocessing and post-processing around memory-bound inference. These heterogeneous kernels exhibit fundamentally different scaling behavior, yet existing desynchronization studies largely attribute performance effects to communication rather than kernel interaction. Unlike the communication-heavy halo-exchange workloads dominating prior desynchronization studies, these applications may communicate little between processes but can exploit controlled desynchronization as a performance lever.

This work addresses the missing case of desynchronization driven by heterogeneous kernel interaction rather than communication or bandwidth variance alone. Specifically, we ask whether kernel heterogeneity itself can become a beneficial source of desynchronization, and under which combinations of memory-to-compute workload fraction and memory-bandwidth saturation, and communication behavior this effect improves application throughput. While prior work demonstrated that desynchronization effects can be induced and exploited~\cite{afzal2022role,AfzalHW:2023:1}, we show that the same bandwidth threshold governs desynchronization in communication-negligible applications where kernel heterogeneity is the driver. An incidental or deliberate stall can keep execution near the memory-bandwidth saturation threshold, turning a phenomenon traditionally viewed as a performance hazard into an optimization opportunity.

\paragraph*{Contributions} We study desynchronization in multi-kernel applications: the MPI Parallel Optical Flow Solver and a tunable, controlled Pisolver-Triad benchmark. Each combines a compute-bound kernel (Ray Tracing or Pi Solver) with a memory-bound kernel (Optical Flow or STREAM Triad) and negligible inter-process communication. Concretely, we:
\begin{itemize}
  \item quantify the performance impact of naturally occurring
  desynchronization relative to a forced lock-step baseline, and identify the
  memory-to-compute workload fraction that maximizes this gain;
  \item show that eliminating MPI communication stalls via asynchronous
  progress can \emph{degrade} performance, by removing an incidental
  reliefing effect that had been reducing memory-bandwidth contention;
  \item analyze the impact of problem size on L3 bandwidth utilization to identify the best sustain desynchronization;
  \item 
  validate the real hardware traces against simulated traces using DisCostiC to characterize where and why the idealized simulation diverges from real hardware behavior.
\end{itemize}
These findings are distilled into a simple design rule: performance is maximized when concurrent memory-bound ranks remain near the hardware's measured bandwidth-saturation threshold, whether regulated by workload ratio, communication stalls, or deliberate noise injection.

\paragraph*{Open-source Data Source} All data required to reproduce this work are available in our performance artifact repository\footnote{\url{https://go-nhr.de/Multi-Kernel-MPI-Analysis}}. The repository is organized by figure numbers, with each directory containing the corresponding raw results, key numeric data used for plotting, runtime logs, and supporting scripts required for reproduction and figure generation. The artifact also includes input files and documentation of the hardware and software environments used for the experiments, enabling fully reproducible results.

\section{Methodology and Applications}
This section derives the theoretical upper bound on desynchronization speedup, experimental applications, and methodology used to isolate kernel-driven desynchronization from other performance effects.

\subsection{Theoretical Upper Bound on Desynchronized Speedup}
\label{sec:upperbound}
Let $P^{t}_{\mathrm{async}}$ and $P^t_{\mathrm{sync}}$ be maximum total performance under natural desynchronization and forced synchronization (barrier overhead excluded from $P^t_{\mathrm{sync}}$). The desynchronization speedup~\cite{AfzalPhDthesis:2026} is: 
\begin{equation}
S_d = \frac{P^{t}_{\mathrm{async}} - P^t_{\mathrm{sync}}}{P^t_{\mathrm{sync}}} \times 100\ [\%],
\label{eq:Sd}
\end{equation}
which isolates the gain from desynchronization itself, excluding effects such as reduced communication. 
$S_d=0\%$ means no gain, $S_d<0\%$ a loss, and $P^{t}_{\mathrm{async}}/P^t_{\mathrm{sync}}\approx1+S_d/100$.
In the dual-kernel program communication is negligible ($t^{\mathrm{comm}}\rightarrow0$); each iteration has compute-bound time $t^{CB}$ and memory-bound time $t^{MB}$, with $t_{\mathrm{sync}}=t^{CB}+t^{MB}$ and memory-to-compute fraction as: 
\begin{equation}
\rho=\frac{t^{MB}}{t^{MB}+t^{CB}}\in[0,1],
\label{eq:rho}
\end{equation}
reported as a percentage (e.g., $\rho=0.55\rightarrow55\%$). Desynchronization overlaps kernels, reducing the effective memory-bound time to $t^{MB\prime}\leq t^{MB}$ while $t^{CB}$ is unchanged. The ideal per-iteration time is therefore
$\max(t^{MB\prime},t^{CB})$, minimized when $t^{MB\prime}=t^{CB}$.
Substituting $t^{MB}=\rho t_{\mathrm{sync}}$ and $t^{CB}=(1-\rho)t_{\mathrm{sync}}$ gives
\begin{equation}
S_d^{\max}(\rho)=100\cdot\frac{\rho}{1-\rho}\ [\%],
\label{eq:Sdmax}
\end{equation}
valid only while $t^{MB\prime}$ can shrink to match $t^{CB}$ without limit.
In practice $t^{MB\prime}$ is bounded: within the bandwidth-saturation region ($c\in[C_{\mathrm{sat}},N]$ active processes per ccNUMA domain), aggregate bandwidth is roughly constant, so $t^{MB\prime}(c)=t^{MB}\cdot c/N$, giving a floor $t^{MB}_{\min}=t^{MB}\cdot C_{\mathrm{sat}}/N$ at $c=C_{\mathrm{sat}}$. The ideal time becomes $\max\!\left(t^{CB},\,t^{MB}\,C_{\mathrm{sat}}/N\right)$, with Eq.~\ref{eq:Sdmax} holding until this floor dominates, at $\rho=\rho^{*}=N/(N+C_{\mathrm{sat}})$. Combining both regimes:
\begin{equation}
S_d^{\max}(\rho)=
\begin{cases}
100\cdot\dfrac{\rho}{1-\rho}, & \rho\leq\rho^{*}\\[6pt]
100\cdot\left(
\dfrac{N}{\rho\, C_{\mathrm{sat}}}
-1
\right), & \rho>\rho^{*}
\end{cases}
\label{eq:Sdmax2}
\end{equation}
continuous at $\rho^*$, where both branches equal $100\cdot N/C_{\mathrm{sat}}$. For our platform ($N=18$, $C_{\mathrm{sat}}=10$; Figs.~\ref{fig:heatmap}b, \ref{fig:pitriad}c), $\rho^{*}\approx64\%$, and the measured best case $\rho=55\%$ ($<\rho^*$) has a theoretical desync speedup limit -- consistent with the observed behavior of speedup peaking near the best-case region and declining toward both the compute- and memory-heavy extremes.

\subsection{Parallel Optical Flow Solver}
The Parallel Optical Flow Solver\footnote{\href{https://github.com/SebastianKuckuk/ray-tracing-optical-flow}{https://github.com/SebastianKuckuk/ray-tracing-optical-flow}} is an MPI application that synthesizes images via ray tracing and then computes their optical flow fields. Each MPI process independently performs a configurable number of repetitions; within each repetition, it alternates between the compute-bound Ray Tracing (RT) kernel and the memory-bound Optical Flow (OF) kernel across several sub-steps. Because the two kernels place demands on different hardware resources and communication between MPI processes is minimal, natural imbalances in kernel execution time cause processes to drift out of lock-step as execution proceeds.

Two application parameters let us tune the balance between the two kernels. The OF image size is governed by the \emph{multigrid level} $m$, with side length $2^{m}-1$; a higher multigrid level increases the memory-bound OF workload. The RT input size is the OF image size scaled by a \emph{supersampling factor} $s$, i.e., $n_x^{RT}=n_x^{OF}\cdot s$ and $n_y^{RT}=n_y^{OF}\cdot s$; a higher supersampling factor increases the compute-bound RT workload. Sweeping these two parameters lets us move the application's execution profile continuously from compute-bound to memory-bound. 

Each MPI process computes the same global image sequence, but ranks are staggered: rank $r$ begins at frame $r\cdot dt\cdot\mathrm{numTimeSteps}$
rather than frame zero, so that at steady state all ranks are working on different, evenly-spaced frames of the same sequence. Because sphere placement in the synthesized scene follows a deliberately non-uniform (``hourglass''-shaped) spatial distribution, different frames contain different numbers of spheres, which is a secondary, load-imbalance-driven source of inter-process timing variation, distinct from the kernel-driven desynchronization that is the focus of this paper. Below, we describe how this secondary effect is neutralized, allowing the two sources of inter-process variation to be studied independently.

\paragraph*{Isolating Kernel-Driven Effects in Flow solver from Load Imbalance}
\label{sec:mitigation}
Before attributing performance changes to kernel-driven desynchronization, we
first isolate and neutralize two independent sources of initial load
imbalance in the application: the non-uniform, hourglass-shaped spatial
distribution of spheres across frames (controlled by \texttt{sinPrefactor}),
and residual differences between the staggered starting frames of different
ranks (controlled by timestep size $dt$). We quantify per-rank workload imbalance via the total number of
double-precision floating-point operations (FLOPs) after a fixed number of
iterations,
\begin{equation}
\label{eq:flop}
\mathrm{FLOPs}
=
\sum_{i\in\{1,2,4,8\}}
i\,F_i,
\end{equation}
where $F_i$ denotes the number of double-precision floating-point instructions
operating on $i$ values simultaneously (scalar, 128-bit, 256-bit, and
512-bit SIMD instructions, respectively). The weighting factor $i$
accounts for the number of double-precision values processed per
instruction. The instruction counts are obtained using
\texttt{likwid-perfctr}. Box plots of per-rank Flop counts
across 100 repetitions show that variability shrinks monotonically as $dt$ is
reduced, confirming that a sufficiently small $dt$ (we use $10^{-5}$) yields
near-identical per-process frame data and hence temporal load balance; setting
\texttt{sinPrefactor}~$=0$ removes the spatial imbalance directly. With both
parameters set accordingly, an ITAC trace of the first iteration shows a lock-step startup across
all MPI processes. All results
use this mitigated configuration, so that any performance change observed
thereafter is attributable to kernel-driven desynchronization rather than
residual workload imbalance.

\begin{table}[t]
\centering
\caption{Pisolver-Triad tuning parameters (single-node)}
\label{tab:benchparams}
\small
\begin{adjustbox}{width=\columnwidth}
\begin{tabular}{>{\columncolor{blue!7}}lrr}
\toprule
\rowcolor{gray!20}
Cases & Pi calculation steps & Triad array elements \\
\midrule
Compute heavy: $\rho=18\%$ & $6.12\times10^{9}$ & $2.2\times10^{8}$ \\
Memory heavy: $\rho=91\%$ & $6.8\times10^{8}$ & $2.0\times10^{9}$ \\
Best case: $\rho=55\%$ & $3.6\times10^{9}$ & $1.36\times10^{9}$ \\
\bottomrule
\end{tabular}
\end{adjustbox}
\end{table}

\subsection{Pisolver-Triad Benchmark}
Because desynchronization in the flow solver application sets in within the first few iterations, it is difficult to study its onset and progression directly.
We therefore use the MPI-parallel Pisolver-Triad benchmark\footref{foot:Fritz}, a lightweight benchmark that reproduces the same dual-kernel structure: a compute-bound
$\pi$-calculation kernel (Pisolver) and a memory-bound STREAM Triad kernel.
Each iteration
follows the same interleaving as the application's RT/OF sub-steps: a
Pisolver step, a non-blocking \texttt{MPI\_Isend} to the successor rank,
three further interleaved Pisolver--Triad sub-steps, a blocking
\texttt{MPI\_Recv} from the predecessor, and a final Triad step. 
As in the application, communication is
periodic and lightweight: each rank sends to its predecessor and receives
from its successor, with the time spent in communication is matched to the application's
best-case configuration (using a message size of (200\,KB). Table~\ref{tab:benchparams} lists the tuned kernel
parameters (per-node; scaled linearly by node count for multi-node runs) for
the three configurations used throughout this paper.
The relative weight and array sizes of each kernel, as well as artificial noise, can be
injected at specified ranks, timesteps, and ccNUMA domains -- control that is
not available in the full application.

\section{Experimental Setup}
This section presents the hardware and software environment, measurement methodology, and simulation framework used throughout the evaluation.

\subsection{Testbed and Experimental Methodology}
Experiments were run on the Fritz and Helma\footnote{\label{foot:Fritz}Fritz (\href{https://doc.nhr.fau.de/clusters/fritz}{https://doc.nhr.fau.de/clusters/fritz}) and Helma (\href{https://doc.nhr.fau.de/clusters/helma}{https://doc.nhr.fau.de/clusters/helma})}. Table~\ref{tab:hw}
summarizes the hardware. On Fritz, five nodes (360 MPI processes) were used for
multi-node experiments, at a fixed clock frequency of 2.4\,GHz with
Hyper-Threading disabled. 
Experiments on Helma used a single node with 384 MPI processes.
The application,
benchmark, and simulator were built with Intel oneAPI (version 2025.2) and Intel MPI (version 2021.16) at \texttt{-O3~-xHost} (Cluster A) and \texttt{-O3~-mavx2~-mfma} (Cluster B). Execution traces were collected with the
Intel Trace Analyzer and Collector (ITAC)~\cite{iwainsky2009comparing} and Vampir~\cite{knupfer2012scorep}; memory bandwidth and floating-point
rates were measured with LIKWID's \texttt{likwid-perfctr}~\cite{treibig2010likwid}, using the
\texttt{MEM\_DP} hardware-event group to compute per-process floating-point
operation counts (Eq.~\ref{eq:flop} in Section~\ref{sec:mitigation}). Intel MPI asynchronous progress was
enabled using \texttt{I\_MPI\_ASYNC\_PROGRESS=1} and
\texttt{I\_MPI\_ASYNC\_PROGRESS\_THREADS=1} with one progress thread per
MPI process. Because MPI asynchronous progress is incompatible with ITAC, the real-side
comparison trace in this experiment was instead captured via manual
high-resolution timestamping (JSON, visualized with \texttt{chrome://tracing})
with asynchronous progress enabled.

High-resolution timings were captured with \texttt{std::chrono} around the
kernels of interest, after discarding warm-up iterations (10 for the
application, 80 for the benchmark, followed by an \texttt{MPI\_Barrier}) to
avoid first-call overhead. Variations characterization
used 100 repetitions per timestep-size
configuration and reported as a full
distribution, not a point estimate.
Benchmark experiments ran for 10{,}000 iterations. Barrier
overhead itself (8.03\,$\mu$s for our 5-node configuration, measured with the Intel MPI Benchmarks) is subtracted from forced-synchronous timings so reported speedups reflect only the effect of desynchronization, not instrumentation cost. 
We report single representative runs per configuration, selected after confirming qualitative consistency across preliminary trials. 
For process timeline plots, during execution, LIKWID monitored all cores on each node using the
\texttt{MEM\_DP} event group with 200\,ms sampling intervals to capture memory
bandwidth and floating-point activity.

\begin{table}[t]
\centering
\caption{Hardware configuration of Fritz and Helma.}
\label{tab:hw}
\small
\begin{adjustbox}{max width=\linewidth}
\begin{tabular}{>{\columncolor{blue!7}}lll}
\toprule
\rowcolor{gray!20}
Component & Fritz\footref{foot:Fritz} & Helma\footref{foot:Fritz}  \\
\midrule
Processor & Intel Xeon Platinum 8360Y & AMD EPYC 9965 \\
Microarchitecture & Ice Lake & Zen5c \\
Cores/node & 72 (dual-socket) & 384 (dual-socket) \\
NUMA domains/node & 4 & 8 \\
L3 cache (shared/socket) & 54\,MiB & 384\,MiB \\
Memory / node & 256\,GiB, 8$\times$DDR4-3200 & 768\,GiB, 24$\times$DDR5-4800 \\
Theor. MEM BW/socket & $204.8$\,GB/s & $460.8$\,GB/s \\
Interconnect & HDR100 InfiniBand, fat tree & NDR InfiniBand \\
\bottomrule
\end{tabular}
\end{adjustbox}
\end{table}
\subsection{DisCostiC Simulation Framework}\label{sec:discstic}
\begin{figure*}[tbh!]
\centering
    \subfloat[Performance: OF and RT kernels]{\includegraphics[width=0.35\linewidth]{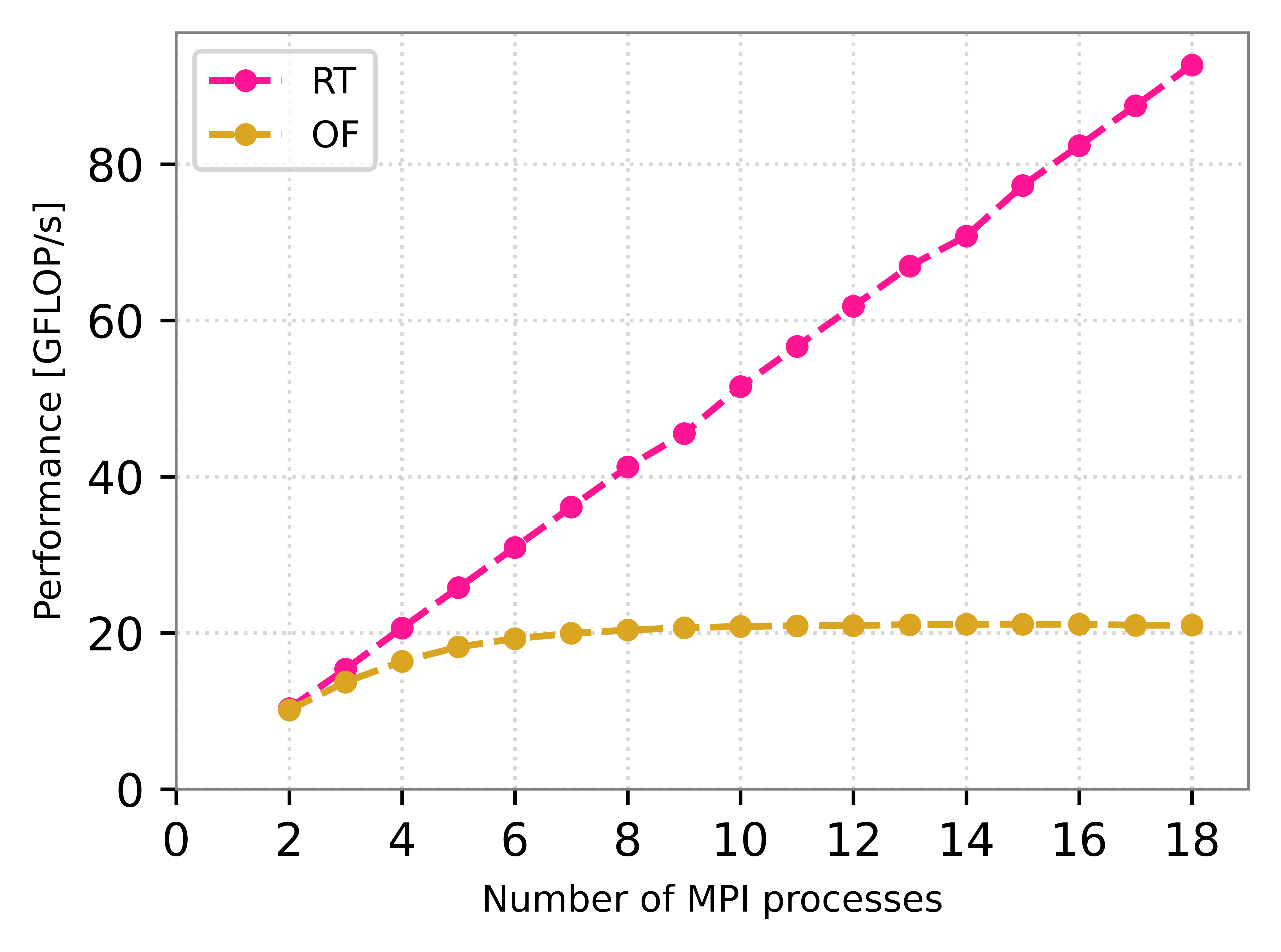}}
    \subfloat[Memory bandwidth: OF kernel]{\includegraphics[width=0.35\linewidth]{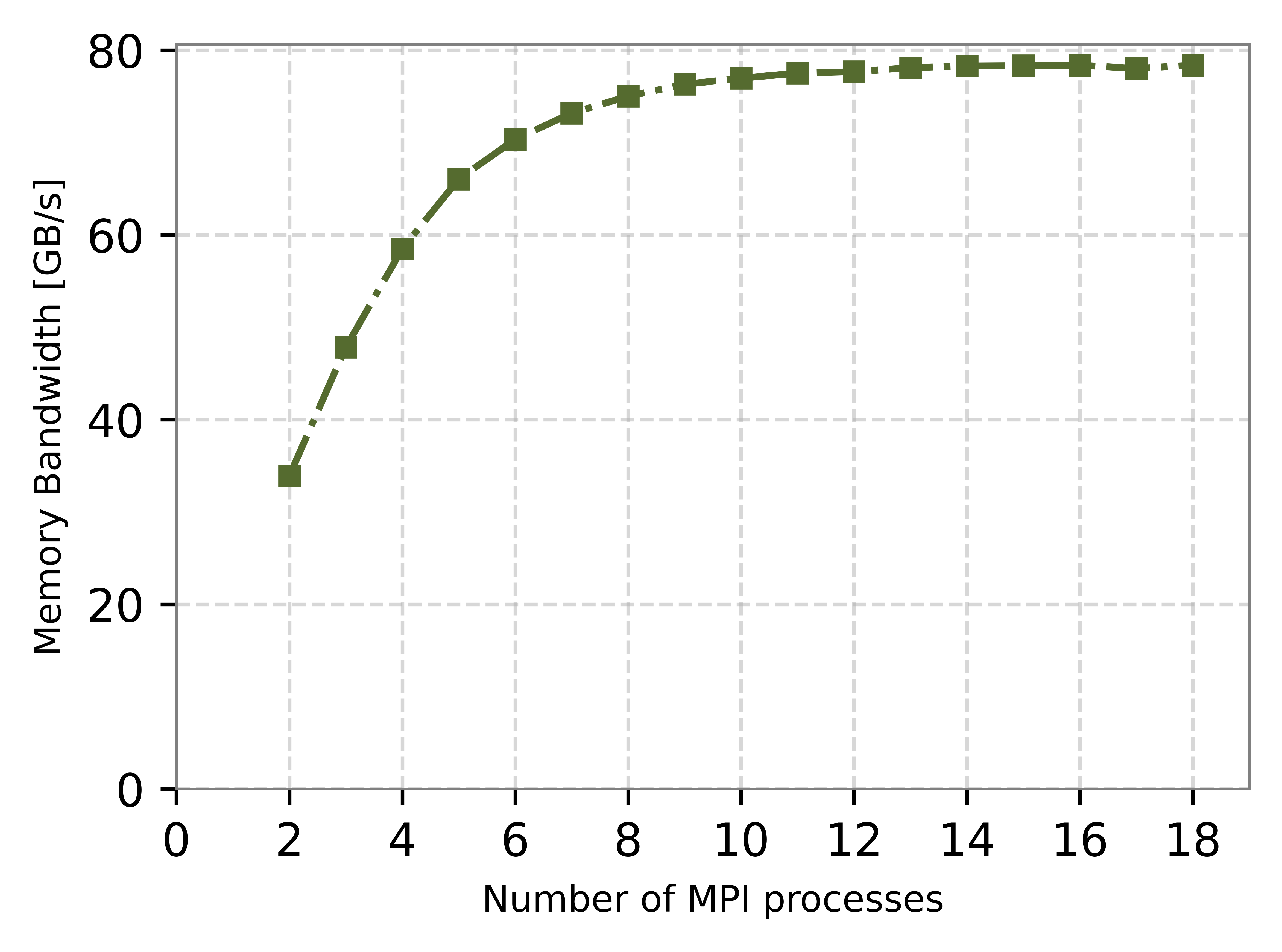}}
    \subfloat[Performance: full application]{\includegraphics[width=0.306\linewidth]{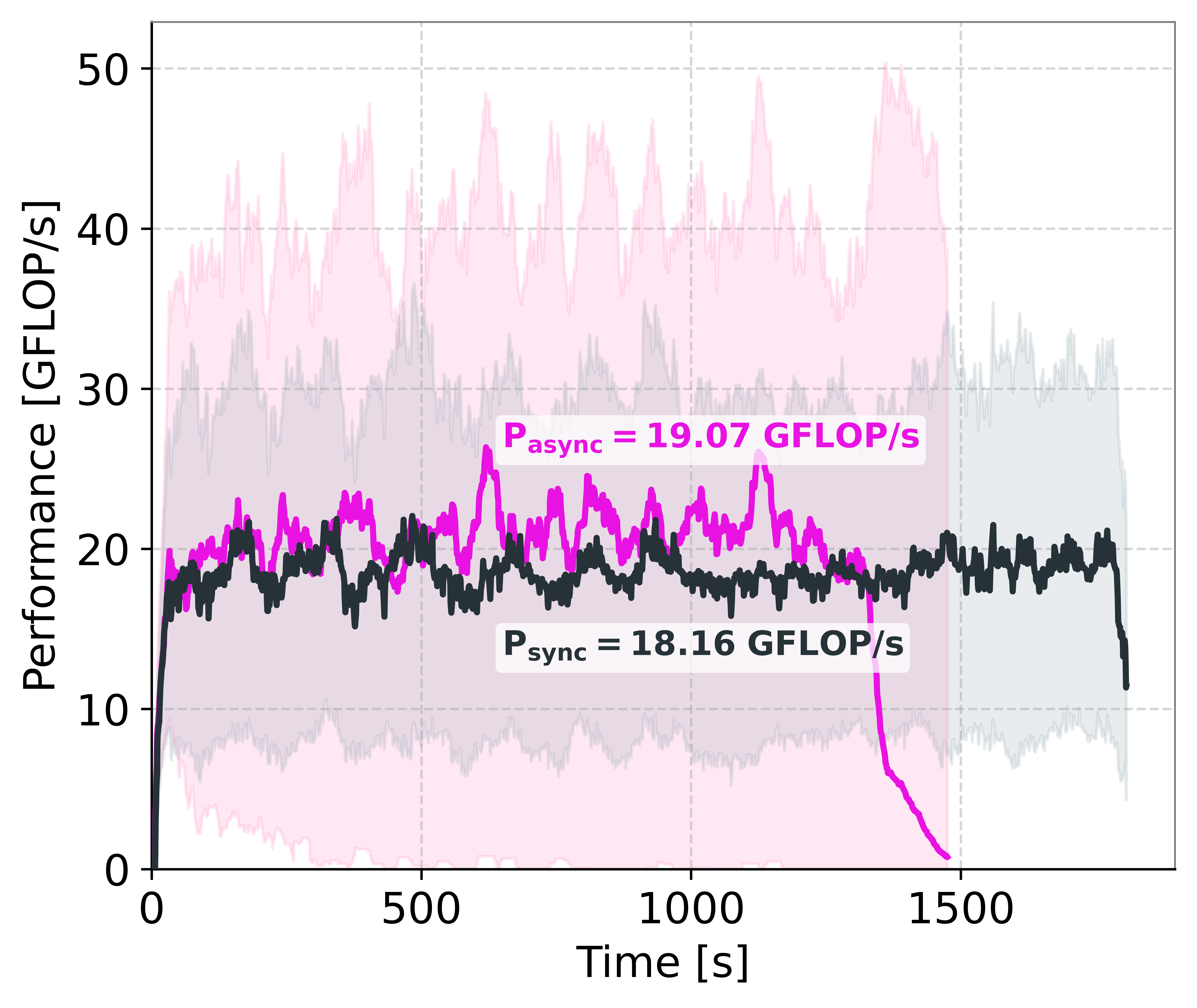}}
\caption{(a) RT and OF kernel performance (GFlop/s) versus active cores within
a ccNUMA domain; RT scales linearly, while OF saturates. (b) OF memory
bandwidth versus active cores, saturating at $\approx$80\,GB/s beyond 7 cores.
(c) Mean per-process performance timeline for the naturally desynchronized
run on five Intel Ice Lake nodes; shaded region shows the min--max range across processes. Performance
increases with desynchronization and stabilizes at $P_{\mathrm{async}}$.}
\label{fig:flowsolver}
\end{figure*}

DisCostiC (Distributed Cost in Clusters) \cite{AfzalPoster:2024,AfzalPoster:2025} is a DSEL-based
simulation framework that predicts MPI application performance using
first-principles analytic models rather than executing application code. Application behavior is
expressed in a Domain-Specific Embedded Language (DSEL) that mirrors the
structure of the original MPI program, the same sequence of compute
kernels and point-to-point communication calls, so a trace can be
generated without re-executing the application itself.
Per-kernel runtimes in DisCostiC are estimated with the Execution-Cache-Memory (ECM)
model~\cite{treibig2009bandwidth}, which
predicts a kernel's cycles-per-cache-line cost as the maximum of the in-core
execution time and the non-overlapping data-transfer time through the cache
and memory hierarchy: $T_{\mathrm{ECM}} = \max(T_{\mathrm{OL}},\,
T_{\mathrm{nOL}}+T_{\mathrm{data}})$. For the compute-bound Pisolver kernel,
each divide operation costs $\approx$ 4 cy/iteration; with 8 double-precision
values per 64-byte cache line, $T_{\mathrm{OL}} = 32$ cy/CL and all
data-transfer terms are negligible, giving 1.667 ns/iteration at 2.4\,GHz.
For the memory-bound Triad kernel, we derive the data-transfer time per cache
line by summing the non-overlapping in-core cost and the transfer times
across each cache level:
\begin{equation}
T_{\mathrm{data}} = T_{\mathrm{nOL}} + T_{L1L2} + T_{L2L3} + T_{L3\mathrm{MEM}}
\approx 22.5\ \mathrm{cy/CL},
\end{equation}
with $T_{L3\mathrm{MEM}} \approx 5.5$~cy/CL derived from a measured attainable
bandwidth of 111.69\,GB/s, consistent with Kerncraft-tool
estimates~\cite{hammer2017kerncraft} and measured timings on Fritz. We scale this cost by the number of cores
concurrently active in a ccNUMA domain to capture bandwidth contention -- the
same saturation effect responsible for the empirically observed 7-core
threshold (Fig.~\ref{fig:flowsolver}b and Fig.~\ref{fig:pitriad}b in later Sections). For the OF kernel, whose per-iteration
cost varies with the recursive multigrid problem size, we fit a logistic
curve to execution time vs.\ problem size and calibrate its scale to match
the measured best-case runtime, then reuse the fitted ECM parameters across
problem sizes.
We support \emph{two-kernel} modeling within DisCostiC: each rank's per-iteration runtime is composed from independent ECM
estimates for both kernels, combined with the bandwidth-contention model, so a rank's simulated progress depends on how many other ranks are simultaneously executing the memory-bound phase -- precisely the coupling that drives desynchronization in the real application. This lets us reproduce and study desynchronization dynamics for both the benchmark and the full application under fully controlled, noise-free conditions.

\begin{figure*}[tbh!]
\centering
    \subfloat[Memory-to-compute fraction $\rho$]{\includegraphics[width=0.48\linewidth]{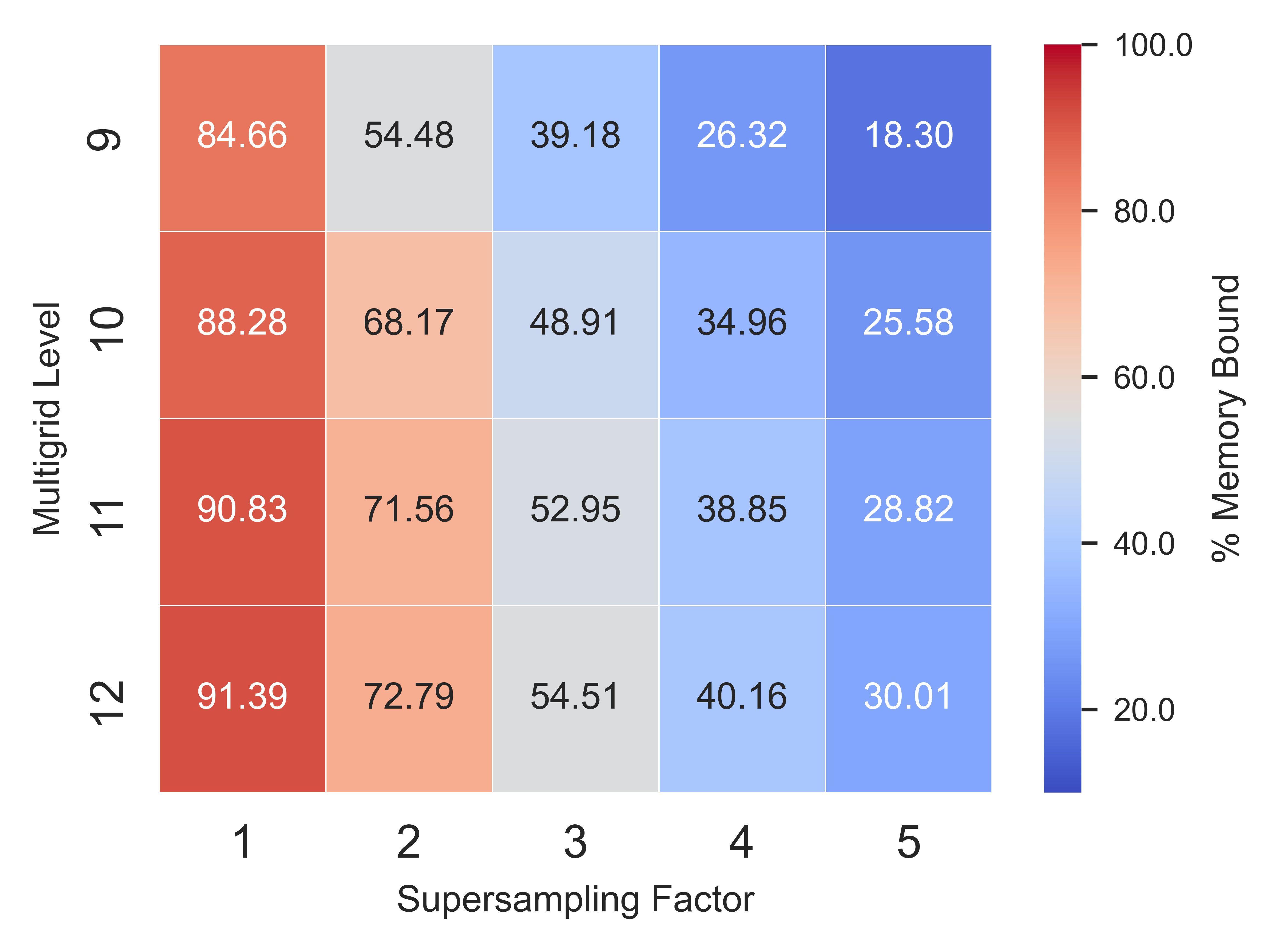}}\quad
    \subfloat[Associated desync speedup $S_d$]{\includegraphics[width=0.48\linewidth]{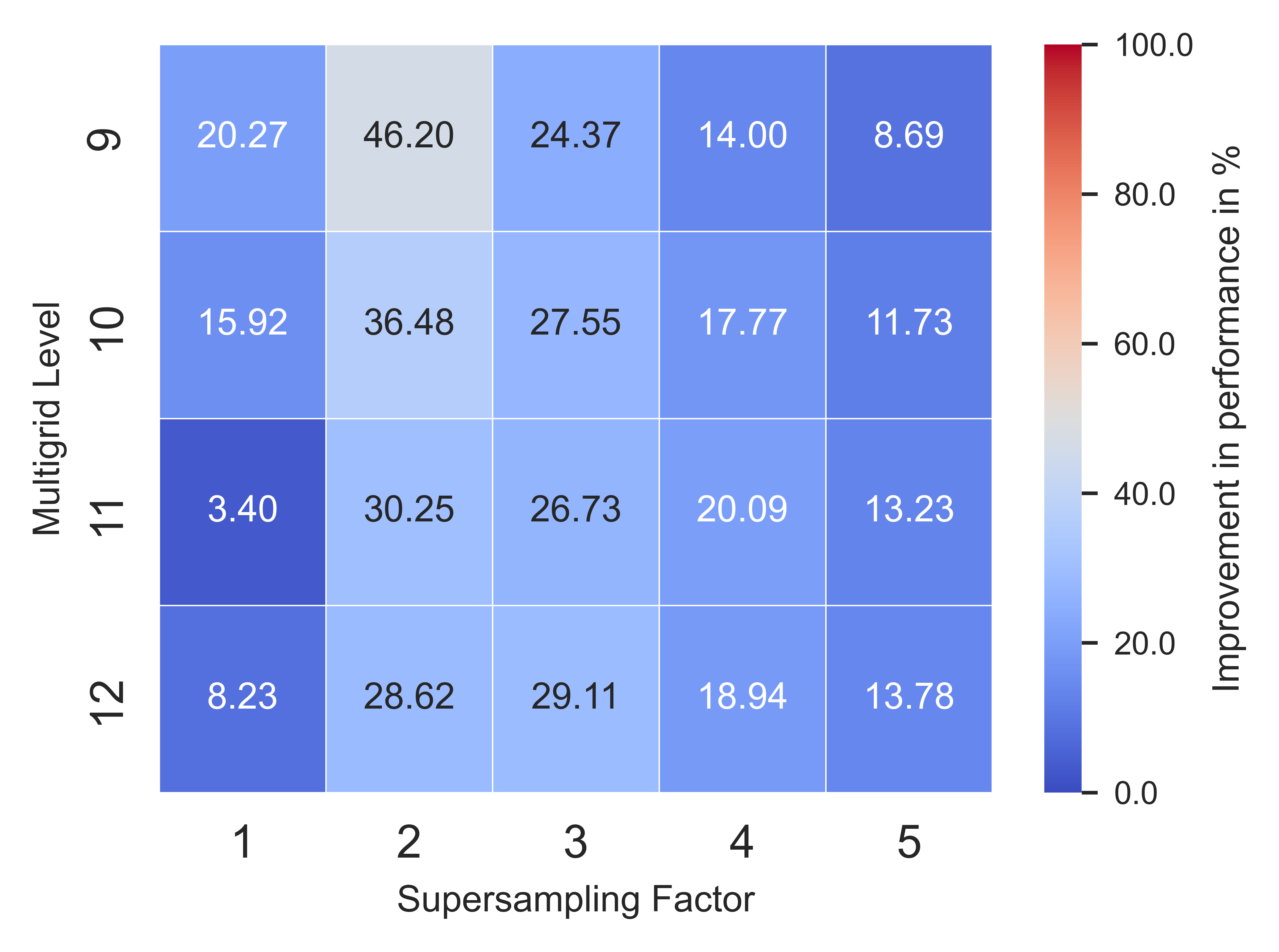}}
\caption{(a) Heatmap of the memory-to-compute fraction $\rho$ (Eq.~\ref{eq:rho} for the flow
solver under lock-step execution across multigrid levels and supersampling
factors. (b) Corresponding normalized desynchronization speedup (Eq.~\ref{eq:Sd}) from end-to-end runtime (maximum MPI
process time) on five Intel Ice Lake nodes; peak speedup ($\approx46\%$)
occurs near $\rho\approx55\%$.}
\label{fig:heatmap}
\end{figure*}

\section{Results: Optical Flow Solver} 
\label{sec:Solver}
We identified the OF kernel's bandwidth-saturation using
\texttt{likwid-perfctr} measurements within a single ccNUMA domain (multigrid level~11, supersampling factor~2, chosen such that the working
set exceeds cache capacity; $dt=10^{-5}$, \texttt{sinPrefactor}~$=0$), defining the \emph{saturation
core count} as the point at which throughput reaches 95\% of its maximum.
Fig.~\ref{fig:flowsolver}(a, b) shows RT scaling consistently with core count (compute-bound)
while OF throughput saturates at 7 active cores per ccNUMA domain, sustaining
$\approx$ 80\,GB/s, confirming OF's memory-bound classification.
With load imbalance mitigated, execution begins in a synchronized, lock-step
state; as execution proceeds, this synchrony erodes naturally because RT and
OF have different bottlenecks. We compare this naturally desynchronized run
against a forced-synchronous variant with a single \texttt{MPI\_Barrier} inserted at the midpoint of each
iteration after two RT/OF sub-steps (a barrier after full-iteration is infeasible, as it deadlocks against the application's four-sub-step
structure). 
Fig.~\ref{fig:flowsolver}(c) shows mean per-process performance (Flops/s) over time for the default run, with a shaded band
showing the spread across MPI processes. 
Performance is low in the first iterations, while processes are still tightly
coupled and OF has not yet benefited from reduced bandwidth contention. As
desynchronization propagates globally, fewer processes execute OF concurrently,
each obtaining a larger share of memory bandwidth, and performance rises to a sustained plateau. In the final iterations, only the few remaining lagging
processes still run OF, each with near-exclusive bandwidth access. Overall, the naturally desynchronized execution achieves
278.5\ GF/s, compared with
221.7\ GF/s for the forced-synchronous baseline, providing a $\approx25.5\%$ performance improvement due to
desynchronization.

\begin{figure*}[t]
\centering
    \subfloat[Compute-heavy: $\rho=18\%$]{\includegraphics[width=0.25\linewidth]{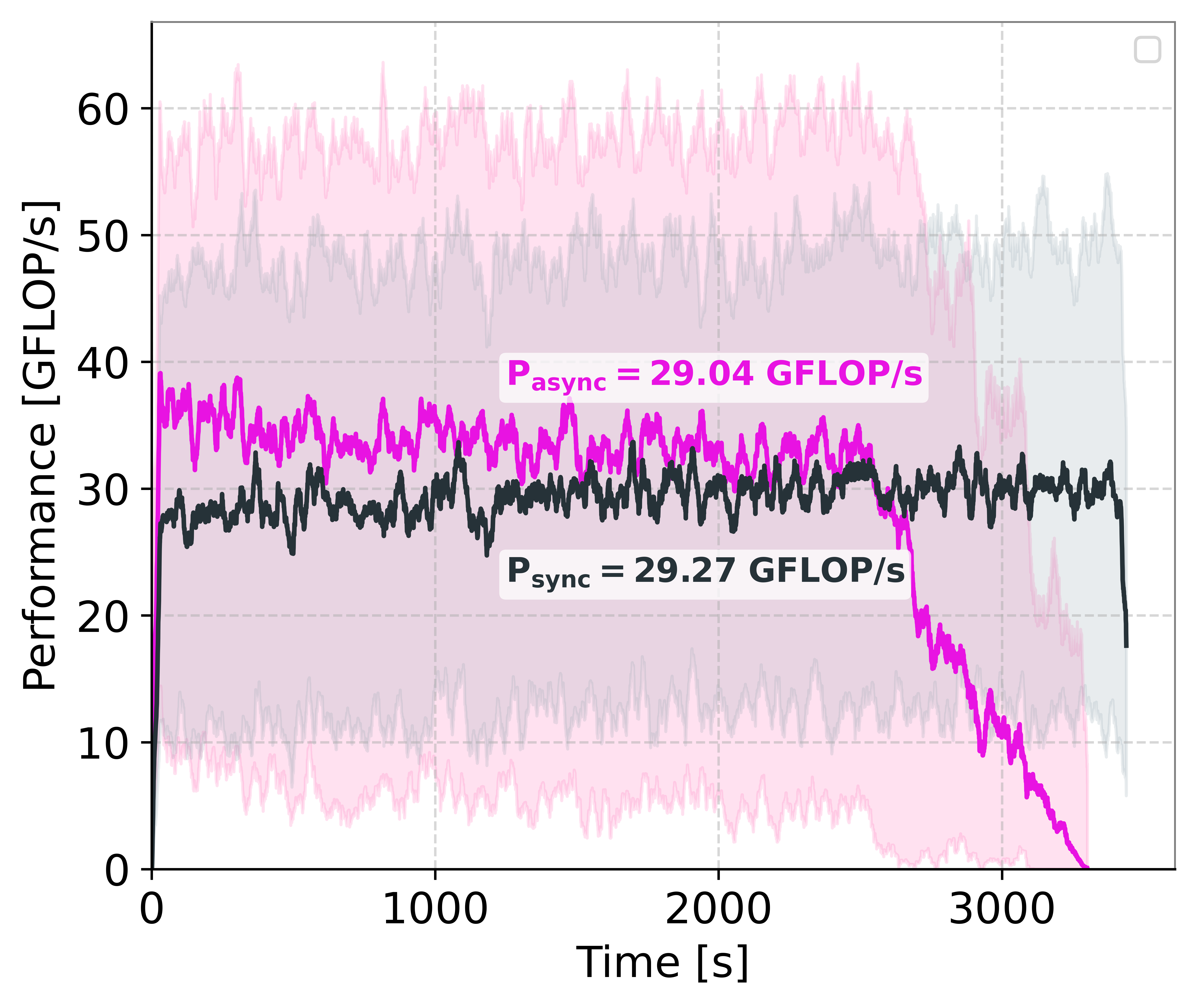}}
    \subfloat[Memory-heavy: $\rho=91\%$]{\includegraphics[width=0.25\linewidth]{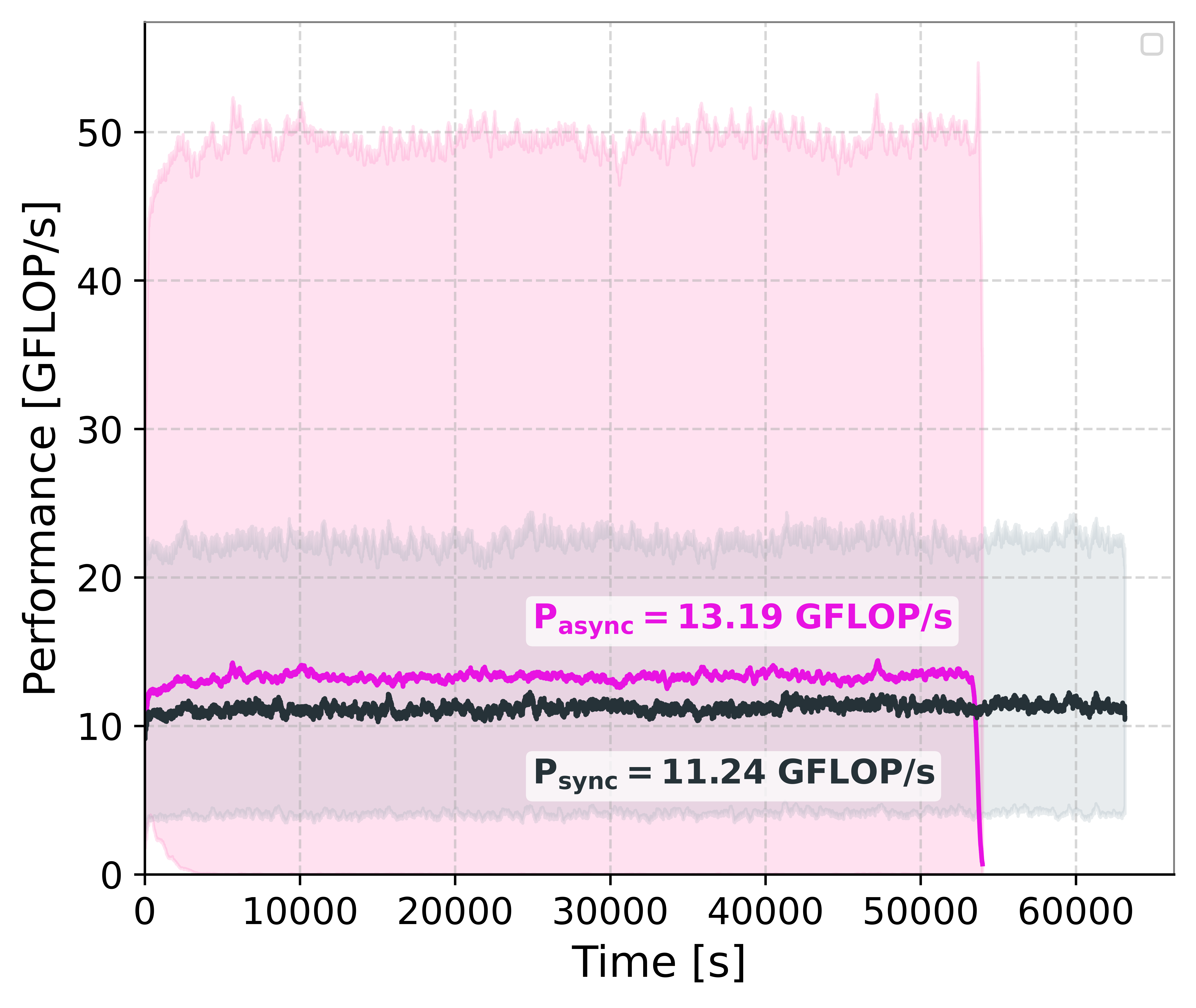}}
    \subfloat[Best case: $\rho=55\%$]{\includegraphics[width=0.25\linewidth]{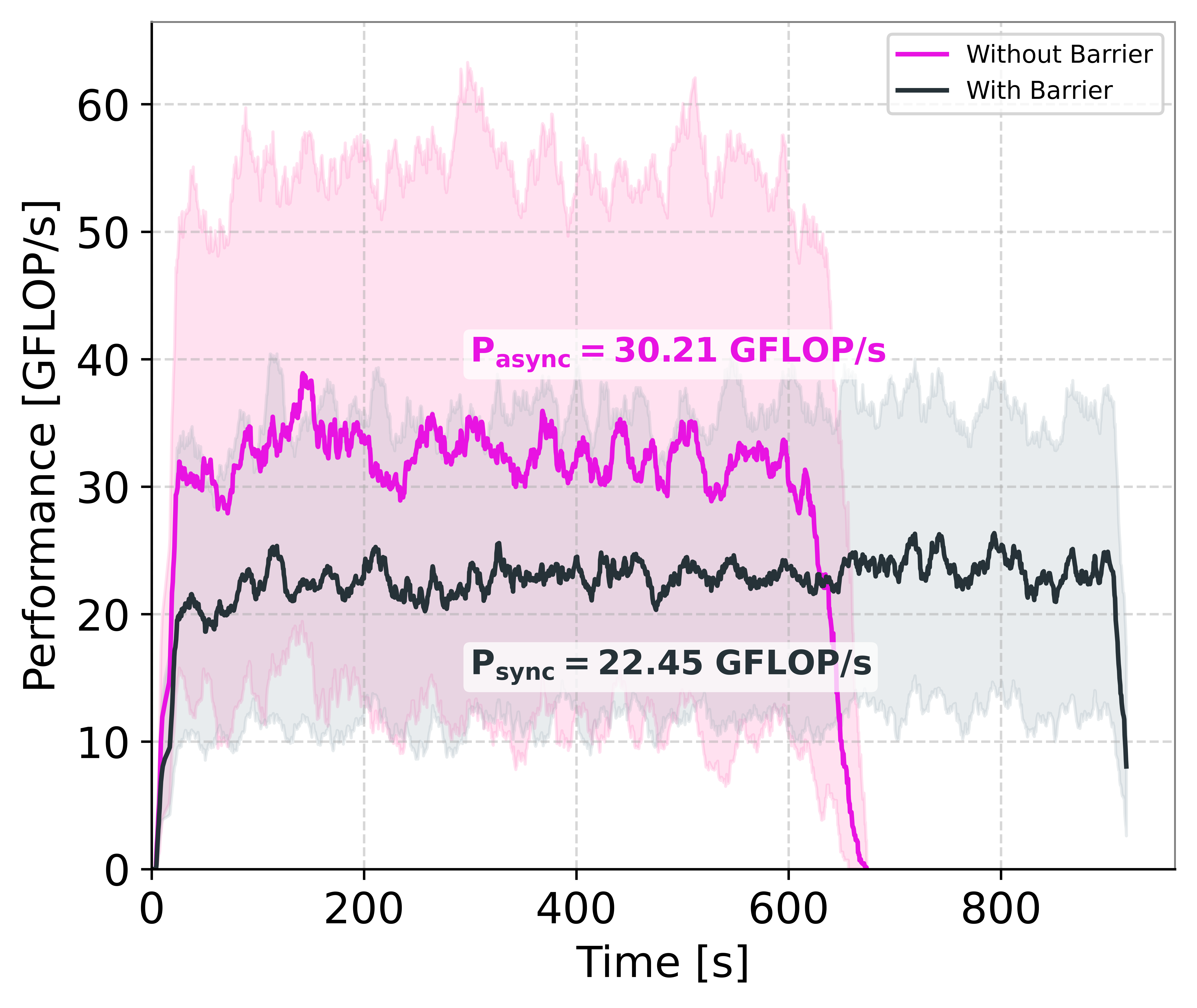}}
    \subfloat[Best case (async progress)]{\includegraphics[width=0.25\linewidth]{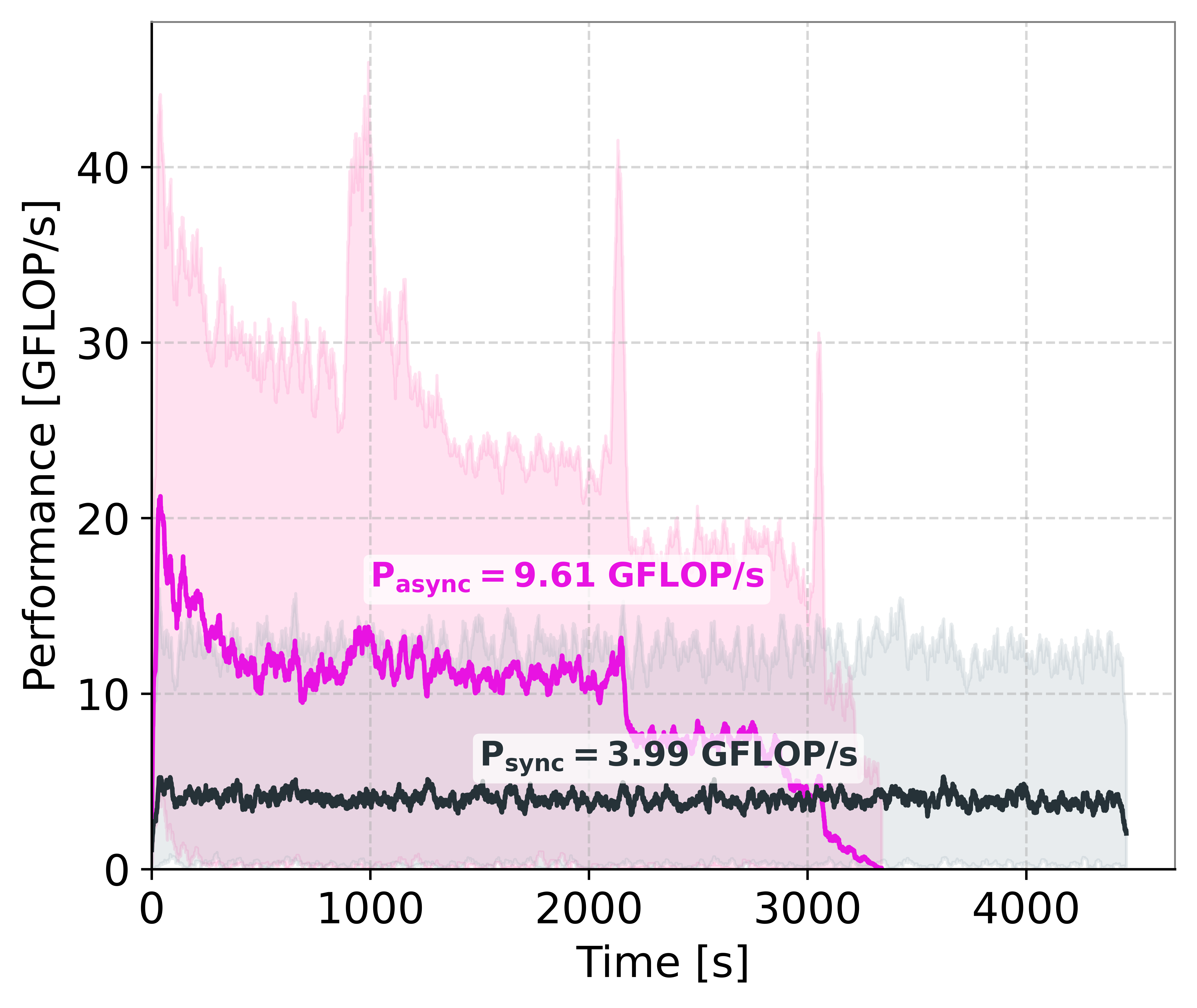}}
\caption{Per-process performance timelines for the flow
solver on five Intel Ice Lake nodes:
(a) compute-heavy $\rho=18\%$, (b) memory-heavy $\rho=91\%$, and best-case
$\rho=55\%$ with (c) disabled and (d) enabled asynchronous progress.
}
\label{fig:ratio}
\end{figure*}

\begin{figure}[t]
\centering
    \subfloat[Best case trace]{\includegraphics[width=0.49\linewidth]{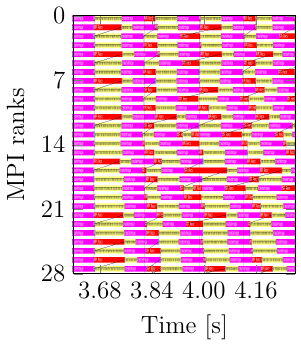}}
    \subfloat[Best case trace (async progress)]{\includegraphics[width=0.49\linewidth]{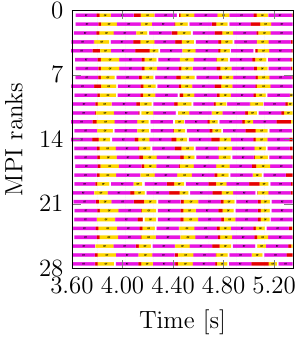}}
    \caption{ITAC traces over four timesteps for the best-case $\rho=55\%$ of the flow
solver on five Intel Ice Lake nodes (a) without and (b) with
asynchronous progress (pink: RT, yellow: OF, red: MPI); long
\texttt{MPI\_Recv} (red) stalls largely disappear with asynchronous progress.}
    \label{fig:ratio2}
\end{figure}

\subsection{Effect of the memory-to-compute fraction}
Sweeping the multigrid level and supersampling factor lets us sweep $\rho$
from compute-heavy to memory-heavy. Fig.~\ref{fig:heatmap}a shows $\rho$ as a
function of the two parameters; Fig.~\ref{fig:heatmap}b shows the
corresponding normalized desynchronization speedup (Eq.~\ref{eq:Sd},
using the first-iteration lock-step performance as the synchronous
baseline). The largest speedup, $\approx46\%$, occurs near the balanced
$\rho=55\%$ configuration (multigrid level~9, supersampling factor~2),
which represents our \emph{best-case} configuration.
Fig.~\ref{fig:ratio}(a)-(c) compares per-process timeline performance for a
compute-heavy ($\rho=18\%$), a memory-heavy ($\rho=91\%$) and the best-case ($\rho=55\%$)
configuration. The compute-heavy case shows uniform per-process performance with
minor iteration-to-iteration variation and little benefit from
desynchronization, since OF barely contributes. The memory-heavy case shows
wider variability but limited net speedup: too many processes saturate bandwidth
simultaneously. The best-case configuration shows the widest per-process spread
and the largest net gain, as processes stagger execution and use memory
bandwidth more efficiently.

To explain \emph{why} the best-case configuration wins, we observe active cores per ccNUMA domain at the execution midpoint and endpoint in each configuration. In the memory-heavy case,
midpoint active-core counts reach 16--18 per domain, more than twice the
7-core saturation threshold, causing sustained over-contention; counts drop only in the final iterations, too late to recover the lost throughput. In the compute-heavy case, active-core counts stay below the threshold throughout, so bandwidth is never the bottleneck, but there is also no memory-bound workload for desynchronization to exploit. The best-case configuration keeps midpoint active-core counts close to the 7-core threshold and drops below it by the final iterations, the only configuration that spends most of its runtime near the saturation boundary from both directions.
We note that desynchronization onset is rapid across all configurations, within approximately five iterations in every case, which limits our ability to resolve the transition from synchronized to desynchronized
behavior in fine temporal detail.
For the best configuration $\rho=55\%$, 
the application achieves less than the theoretical ceiling (Eq.~\ref{eq:Sdmax}), with the remaining gap due to imperfect process staggering, boundary-rank effects, and residual serialization discussed in Section~\ref{sec:asyncprog}.

\subsection{The Cost of Removing Communication Stalls}
\label{sec:asyncprog}
Trace analysis of the best-case configuration (Fig.~\ref{fig:ratio2}(a)) revealed that, despite using
non-blocking \texttt{MPI\_Isend}, several processes still block at
\texttt{MPI\_Recv} in early iterations -- a side effect of the MPI
eager/rendezvous protocol switch for messages exceeding the eager threshold,
under which a nominally non-blocking send behaves quasi-blockingly while it
waits for receiver readiness. These stalls behave like idle waves\cite{markidis2015idle} that
propagate into later iterations, in some cases persisting for nearly the
duration of an entire OF sub-step. We eliminated them by enabling MPI
asynchronous progress, offloading communication to a
background thread.
At a single timestep, removing the stalls only eliminates idle time as expected.
However, over longer execution with a larger timestep size (4),
enabling asynchronous progress \emph{reduces} performance by nearly 80\%;
see Fig.~\ref{fig:ratio}d and~\ref{fig:ratio2}b). 
With the stalls present, only a limited
subset of processes execute OF at a time, and those processes benefit from higher
per-core bandwidth; removing the stalls lets all processes make uniform forward
progress, but this increases the number of processes concurrently competing for
memory bandwidth, pushing the system well past the 7-core saturation
threshold and reducing per-process throughput. In this workload, the natural
partial serialization induced by communication stalls was, in effect, an
unintentional but beneficial throttling mechanism, the same mechanism responsible for the gains, now removed.

\subsection{Cross-Architecture Validation on AMD CPU}
\label{sec:amdvalidation}
\begin{figure}[t]
\centering
    \subfloat[Per-process performance]{\includegraphics[width=0.5\columnwidth]{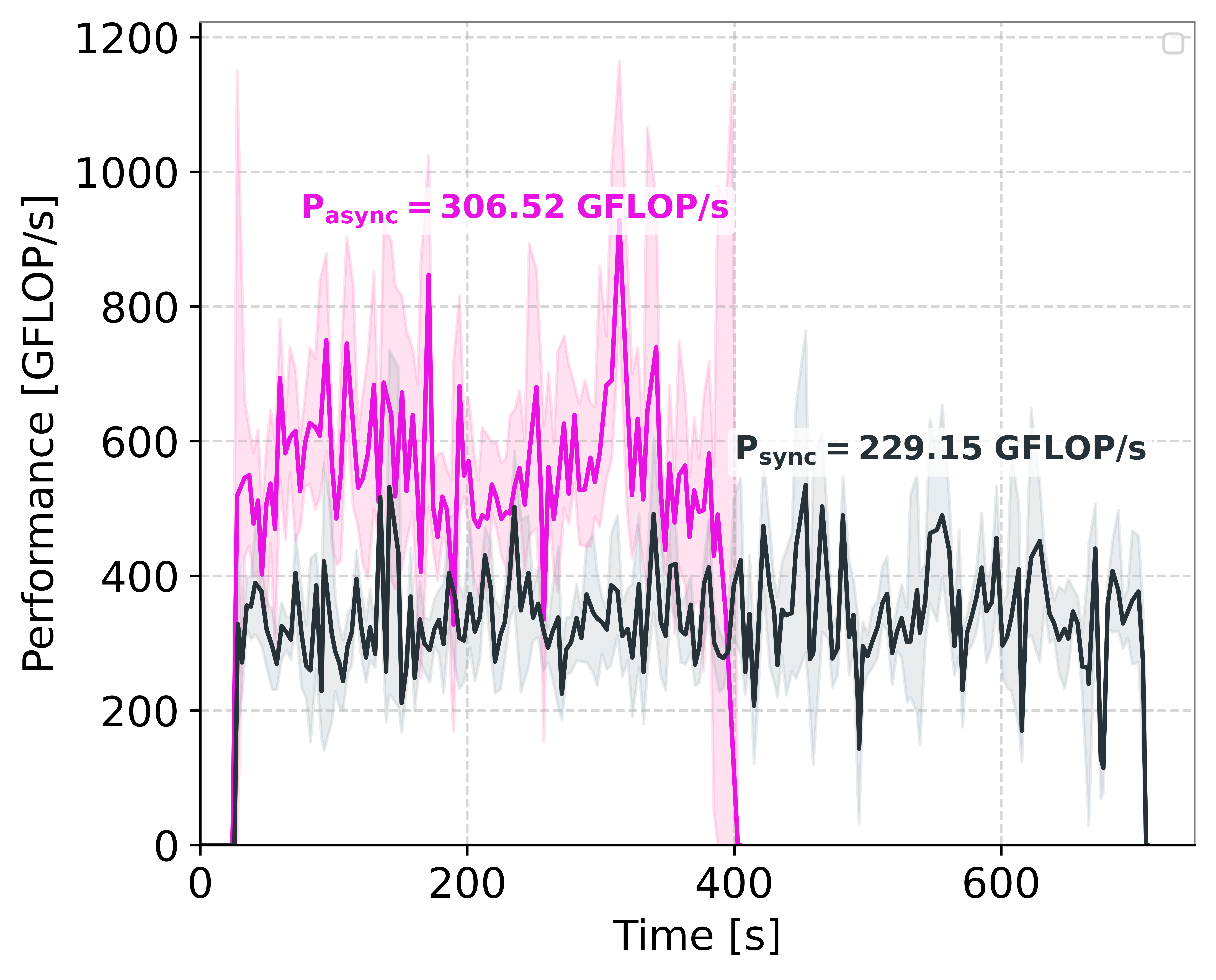}}
    \subfloat[Process timeline]{\includegraphics[width=0.48\columnwidth]{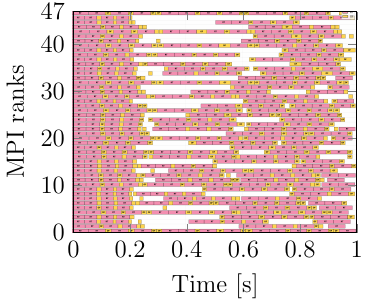}}
    \caption{Best-case memory-to-compute fraction ($\rho=55\%$) on a single AMD
    CPU node (384 MPI processes) of Helma. (a) Per-process performance
    timeline. (b) Initial 1\,s execution trace of one NUMA domain (48 MPI
    processes); the full-node trace is available at
    \url{https://go-nhr.de/AMDnode}.}
    \label{fig:helma}
\end{figure}

To evaluate whether the observed desynchronization benefits are specific to
the Intel Ice Lake platform, we repeated the best-case experiment on Helma,
an AMD EPYC system with a different microarchitecture, memory subsystem, and
NUMA topology. Using the same memory-to-compute balance
($\rho=55\%$), the Optical Flow Solver achieves an $88\%$ performance
improvement from natural desynchronization compared with the forced
lock-step execution. Fig.~\ref{fig:helma}(a) shows the corresponding
per-process performance timeline, where performance increases as processes
gradually desynchronize and memory-bound phases overlap less frequently.
The execution trace of one NUMA domain (Fig.~\ref{fig:helma}(b)) shows the same
qualitative behavior observed on Intel: desynchronization reduces concurrent
memory-bandwidth pressure and allows remaining OF kernels to execute with
higher throughput.
Although the absolute speedup differs from the Intel system due to differences
in core count, NUMA organization, and memory-bandwidth characteristics, the
same mechanism governs the performance gain. The result demonstrates that
beneficial desynchronization is not a hardware- or vendor-specific artifact,
but a general effect arising from the interaction between heterogeneous kernel
scaling behavior and shared memory-bandwidth saturation.

\begin{figure*}[t]
    \centering
    \subfloat[Performance: Pi and Triad kernels]{\includegraphics[width=0.3\linewidth]{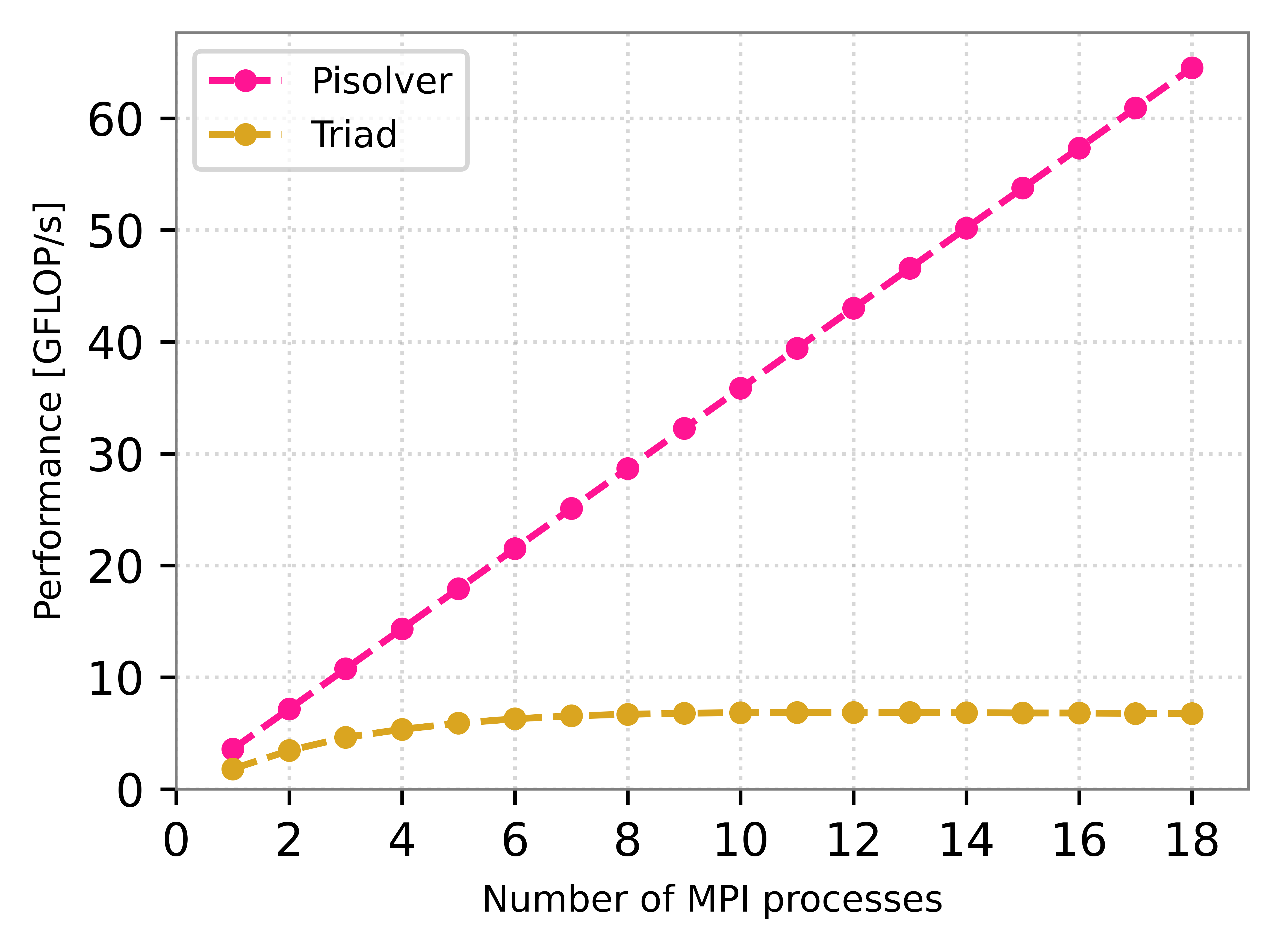}}
    \subfloat[Memory bndwidth: Triad kernel]{\includegraphics[width=0.3\linewidth]{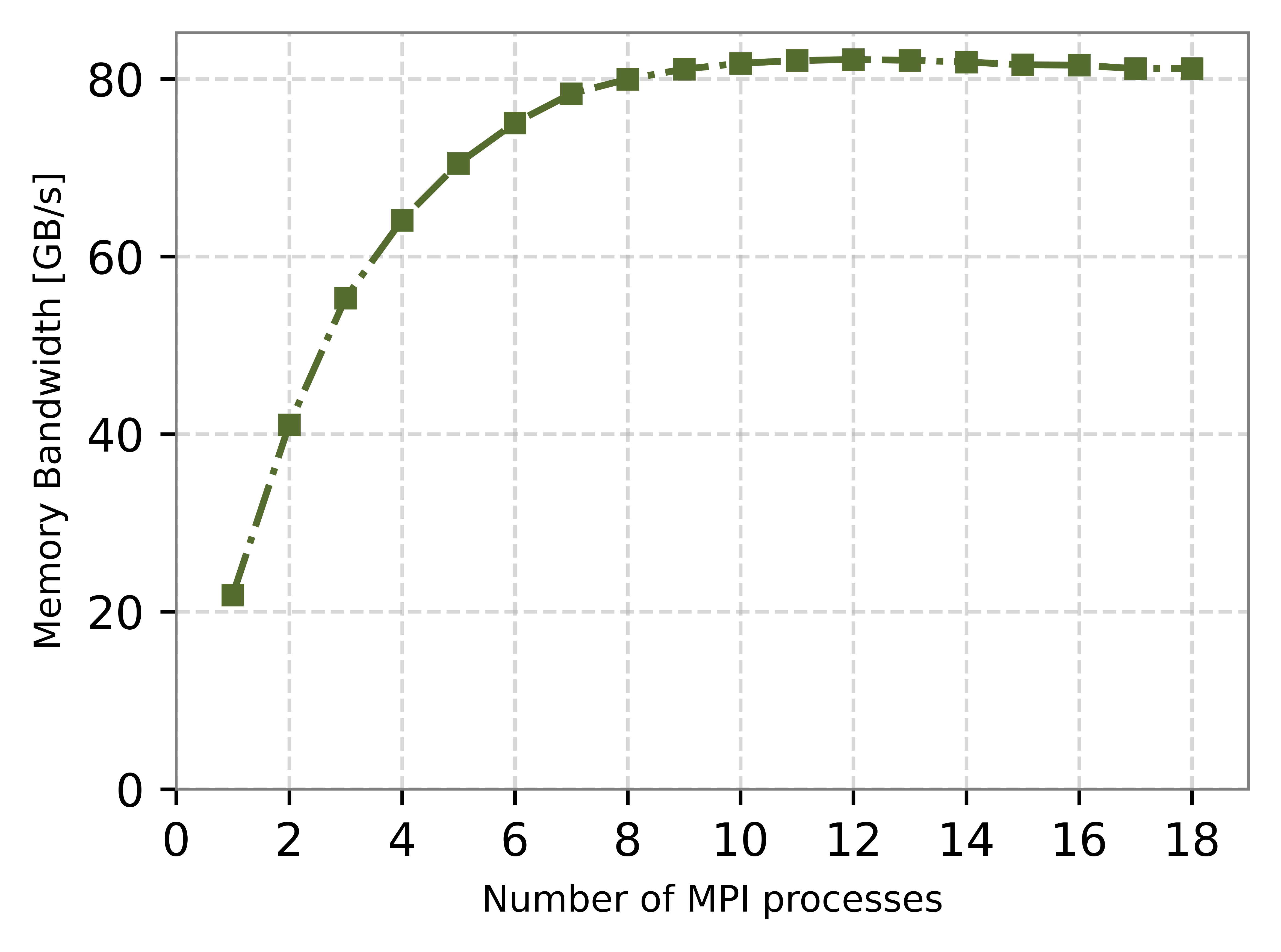}}
    \subfloat[Desync Speedup: Full benchmark]{\includegraphics[width=0.395\linewidth]{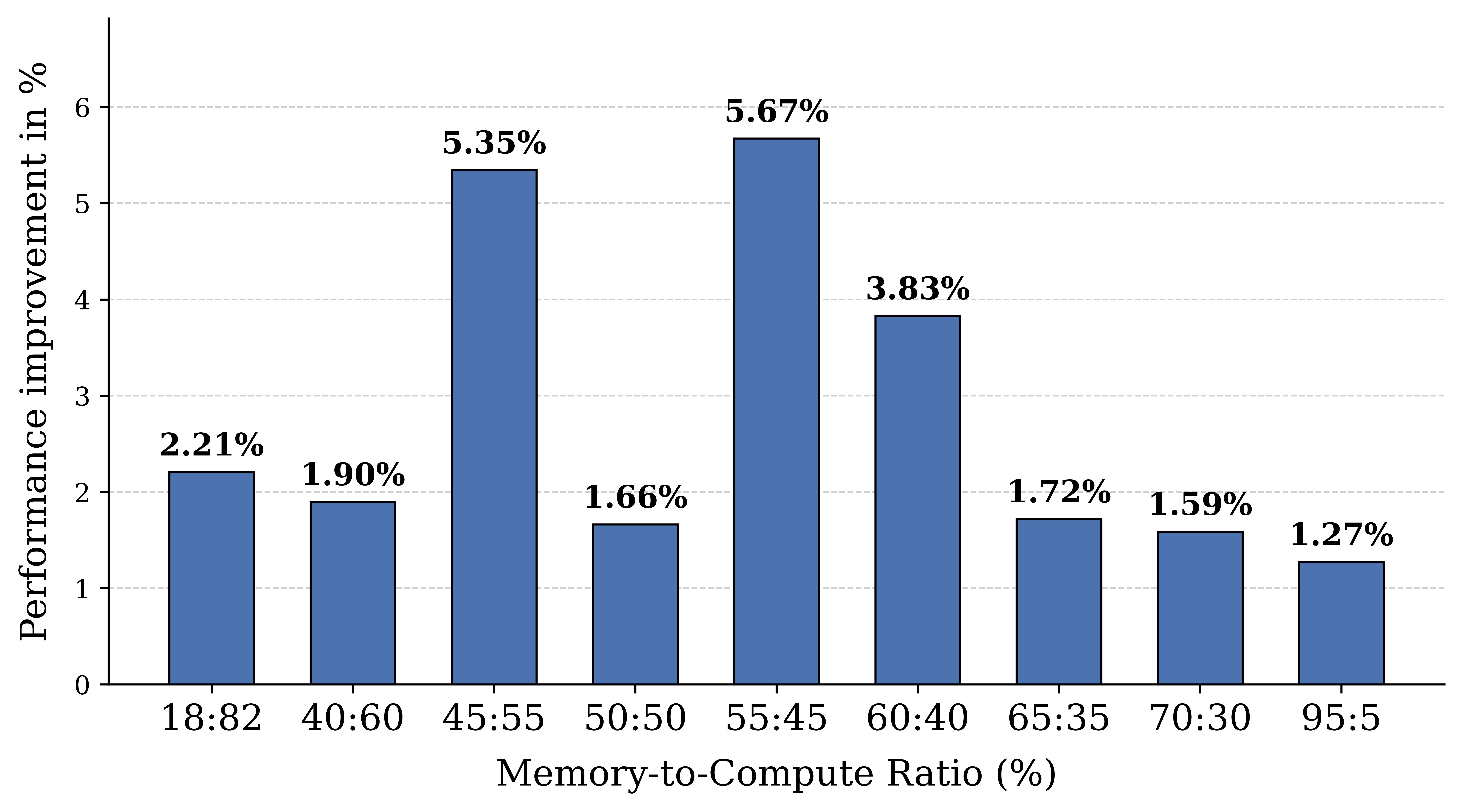}}
    \caption{PI-Triad bench: (a) Performance (GFlop/s) of the Pi and STREAM Triad kernels vs. active cores
within a single ccNUMA domain; Pi scales linearly, STREAM saturates. (b) STREAM memory
bandwidth vs. active cores, saturating at $\approx$80\,GB/s beyond 7 cores.  (c) Corresponding normalized speedup from desynchronization (end-to-end performance with maximum time among all the MPI processes). Peak speedup ($\approx$5.7\%) occurs near $\rho\approx 55\%$.}
    \label{fig:pitriad}
\end{figure*}
\subsection{Simulation Validation with DisCostiC}
\begin{figure}[t]
\centering
    \subfloat[Real trace]{\includegraphics[width=0.5\columnwidth]{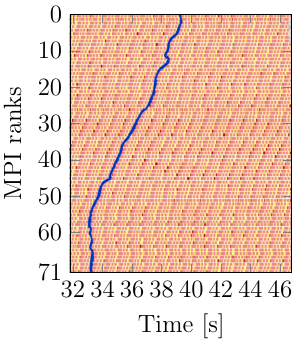}}
    \subfloat[Simulated DisCostiC trace]{\includegraphics[width=0.5\columnwidth]{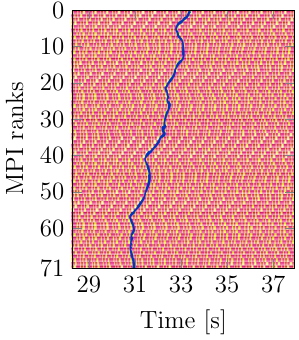}}
\caption{Execution traces of the Optical Flow Solver for the best-case
configuration (29--37\,s) on five Intel Ice Lake nodes: (a) real trace with
asynchronous progress enabled and (b) simulated DisCostiC trace. One selected
iteration completion time is highlighted in blue.}
    \label{fig:simapp}
\end{figure}

Using the ECM-derived parameters from Section~\ref{sec:discstic}, we configured
the two-kernel DisCostiC simulator to reproduce the full solver flow application, and compared simulated traces against real hardware
traces. We replicated the best-case configuration in
DisCostiC using RT/OF kernel timings matched to the real execution, so that
divergence reflects system-level rather than computational differences. We
enabled asynchronous progress to match the noise-free, stall-free conditions
assumed by the simulator.
Both simulated and real traces exhibit early desynchronization and comparable
iteration completion trends (blue line in Fig.~\ref{fig:simapp}). However,
the real trace additionally reflects boundary-rank effects: DisCostiC's DSEL
representation assumes a periodic communication topology, whereas the real
implementation skips \texttt{MPI\_Recv}/\texttt{MPI\_Isend} at domain-edge ranks,
introducing idle stalls or premature progress that are not captured by the
idealized simulation. Despite these differences, DisCostiC reproduces the desynchronization behavior and timing trends of the full application within $<3\%$ difference, validating its use for controlled and repeatable
studies of this phenomenon. The results also identify boundary-process modeling as
a concrete direction for improving simulation fidelity.

\section{Results: Pisolver-Triad Benchmark}
\label{sec:benchmark}
\begin{figure}[t]
\centering
    \subfloat[Real trace]{\includegraphics[width=0.5\columnwidth]{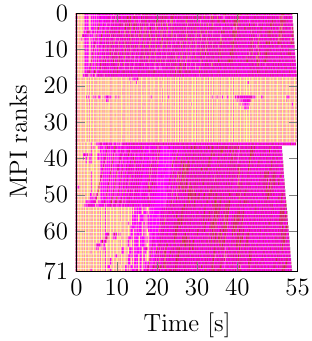}}
    \subfloat[Simulated DisCostiC trace]{\includegraphics[width=0.5\columnwidth]{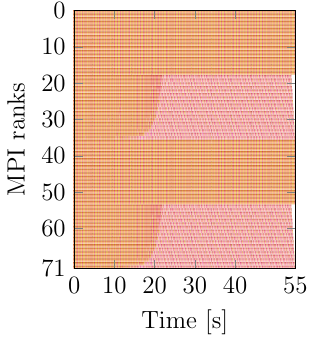}}
\caption{a) Real and (b) simulated DisCostiC traces for the best-case configuration of Pisolver-Triad Benchmark on two Intel Ice Lake chips.}
\label{fig:simbench}
\end{figure}

We first characterize the benchmark's \emph{unperturbed} behavior across the
three tuned configurations (Table~\ref{tab:benchparams}) on a single ICX node
for 10{,}000 iterations. As in the application, Fig.~\ref{fig:pitriad}(a,b)
shows Pi scaling with core count (compute-bound), while Triad saturates at
7 active cores per ccNUMA domain with $\approx$80\,GB/s bandwidth.
Unlike the application, the benchmark shows negligible spontaneous
desynchronization benefit (at most 5.7\%): per-process performance remains
nearly constant over 10k iterations (Fig.~\ref{fig:pitriad}(c)). The sweep of
nine memory-to-compute ratios ($\rho\approx18\%$--$95\%$) using end-to-end
normalized speedup (desynchronized vs.\ forced-synchronous) shows the same
ratio-dependent trend, with a peak of $\approx$5.7\% near $\rho\approx55\%$.
However, the gain is much smaller than in the full application (5.7\% vs.\
46\%) because persistent synchronization in at least one ccNUMA domain limits
the benefit.
Trace inspection confirms that one ccNUMA domain remains synchronized
throughout execution for every configuration (Fig.~\ref{fig:simbench}a),
although the affected domain varies across runs. The changing message
activity from compute- to memory-heavy cases confirms that the tuning controls
the intended workload balance, but additional desynchronization alone does
not improve throughput while a synchronized domain remains the bottleneck.
This motivates deliberate intervention.

\begin{figure}[t]
\centering
    \subfloat[ $k{=}4$]{\includegraphics[width=0.51\columnwidth]{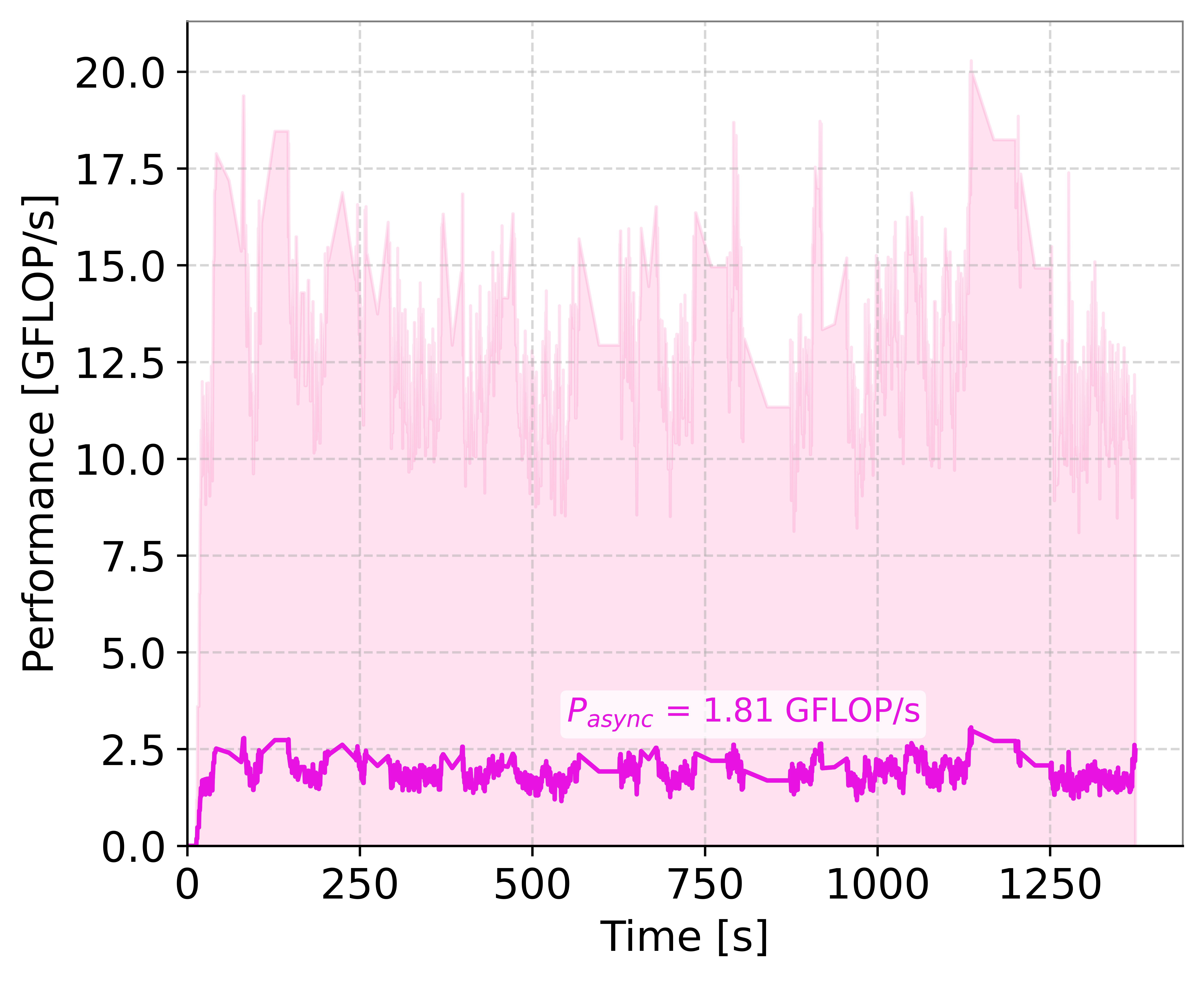}}  
    \subfloat[ $k{=}2$]{\includegraphics[width=0.495\columnwidth]{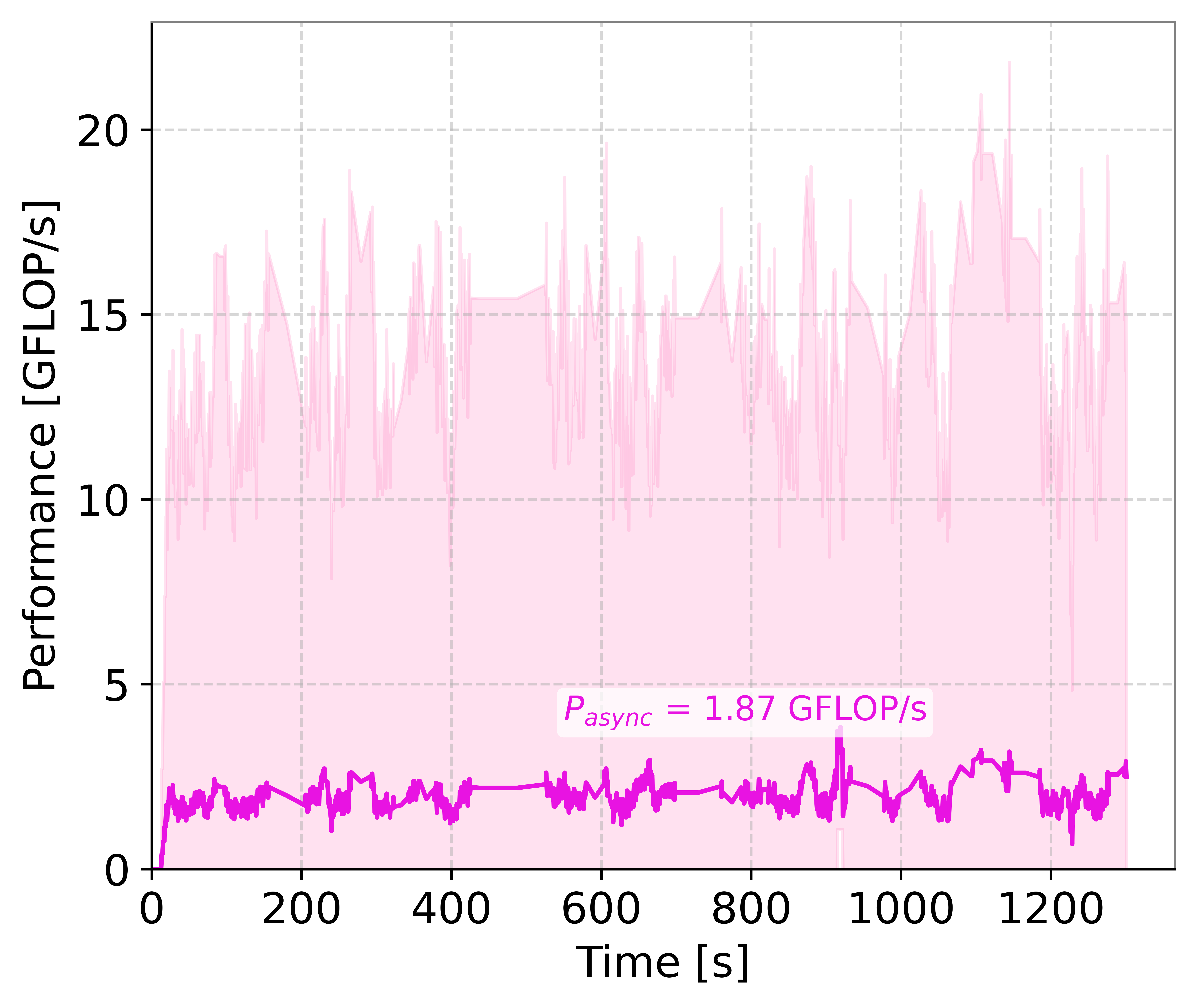}}
\caption{Per-process performance timelines for the best-case configuration of Pisolver-Triad Benchmark with one perturbed process per ccNUMA domain at noise injection frequencies (a) $k{=}4$ and (b) $k{=}2$.}
\label{fig:idlewave}
\end{figure}

\subsection{Periodic, Spatially Distributed Noise Injection}
We first tested whether a single localized delay could break persistent
synchrony by injecting an additional 2-second $\pi$ calculation at one process.
The disturbance propagates through neighbors but is absorbed faster in
memory-heavy cases due to bandwidth contention. All configurations
re-synchronize, showing that isolated delays are insufficient.
We then injected distributed noise: at every $k$th timestep, a random process
performs an additional Pisolver iteration. We evaluated three spatial
granularities: one perturbed process per five-node group, per node, and per
ccNUMA domain. Group-level noise dissipates too quickly, while per-node noise
leaves intra-domain synchrony. Only per-ccNUMA-domain injection provides full
spatial coverage and removes the persistent synchronized domain.
With one perturbed process per ccNUMA domain, noise every fourth timestep
improves performance by 20\% (Fig.~\ref{fig:idlewave}a), while every second timestep increases the gain to
25\%(Fig.~\ref{fig:idlewave}b). Fig.~\ref{fig:idlewave} shows the corresponding performance timelines:
higher injection frequency produces stronger and more sustained
desynchronization. Thus, both spatial coverage and injection frequency are
required to sustain beneficial desynchronization in workloads that do not
naturally desynchronize.

\subsection{Simulation Validation with DisCostiC}
For the unperturbed benchmark (single ICX node, 300 iterations, best-case
configuration $\rho=55\%$), DisCostiC reproduces the baseline execution pattern
and message-exchange behavior observed in the real trace
(Fig.~\ref{fig:simbench}). The main difference lies in the desynchronization
onset: the real execution diverges earlier, whereas the noise-free simulation
develops divergence gradually due to differences in intra- and inter-NUMA-domain
communication latency. Nevertheless, the simulation captures the key behavior
of the real system, with at least one ccNUMA domain remaining internally
synchronized, while predicting runtime with only a $0.5\%$ error.

\begin{figure}[t]
\centering
\includegraphics[width=\columnwidth]{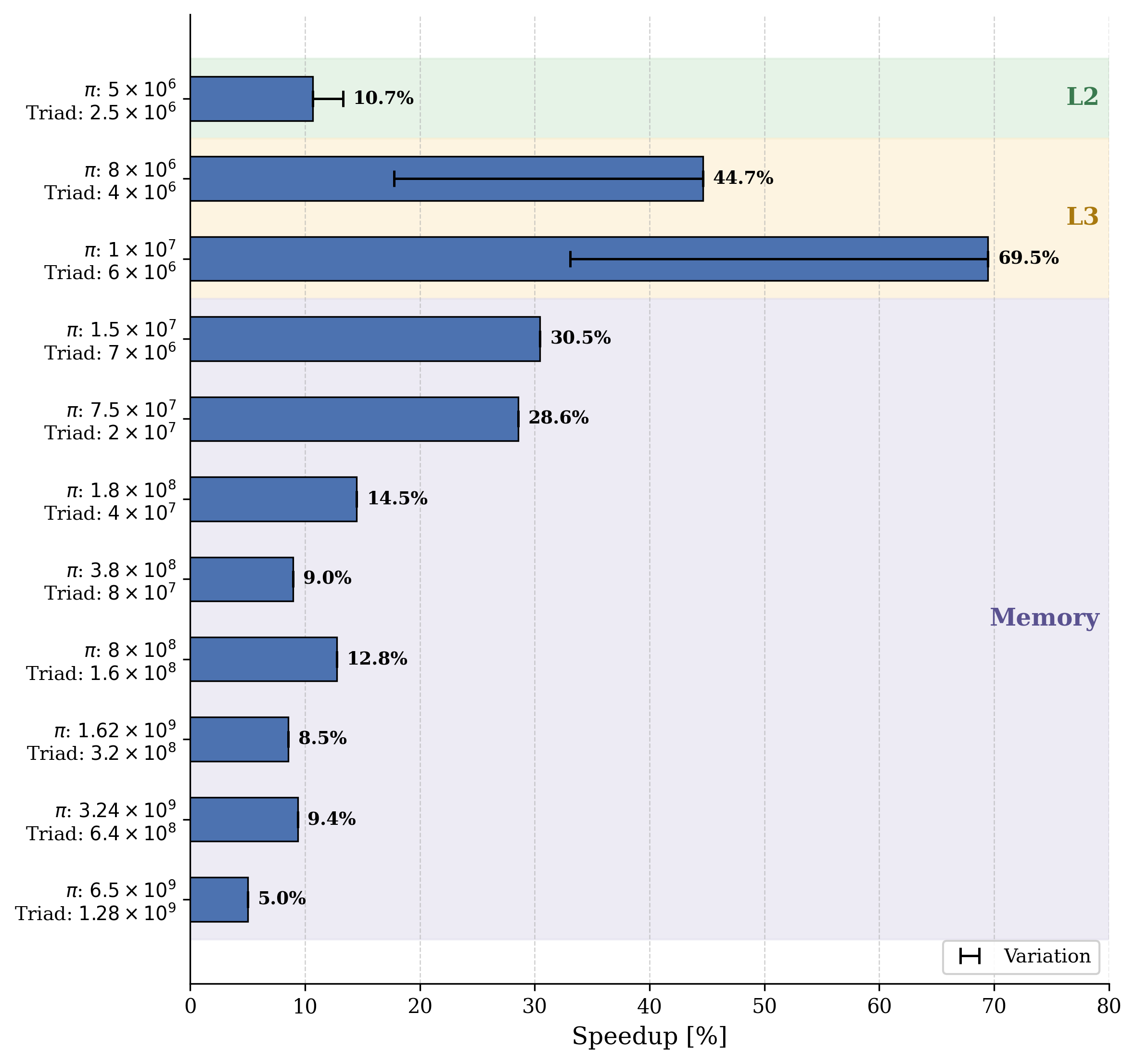}
\caption{Natural desynchronization speedup for the best-case workload fraction
($\rho\approx55\%$) as the per-process Triad array size spans L2-, L3-, and
main-memory-resident working sets on five ICX nodes over 70{,}000 iterations.
For comparison, figures~\ref{fig:pitriad} and \ref{fig:simbench} use a main-memory-resident array
($1.36\times10^{9}$ elements), yielding a 5.7\% speedup. Here, $n$ denotes the
number of Pi calculation steps.}
\label{fig:L3}
\end{figure}
\subsection{Best Case: Rapid Desynchronization When Data Fits in L3}
Previously, the benchmark showed negligible spontaneous speedup ($\leq5.7\%$,
rather than the 20--63\% seen elsewhere) because at least one ccNUMA domain
remains synchronized when the Triad working set resides in main memory,
regardless of workload fraction (Fig.~\ref{fig:pitriad}(c)). We therefore tested
whether this persistent synchrony is a property of main-memory accesses rather
than cache-resident execution.
We fixed the memory-to-compute fraction at the best-case value
($\rho\approx55\%$, Table~\ref{tab:benchparams}) and varied only the absolute
problem size by scaling the Pisolver iteration count and per-rank Triad array
size together, preserving $\rho$. Eleven configurations span Triad arrays from
$2.5\times10^{6}$ to $1.28\times10^{9}$ elements, covering per-domain L2, L3,
and main-memory working sets (Fig.~\ref{fig:L3}). To obtain sufficient runtime,
the benchmark was extended to 70{,}000 iterations.
Figure~\ref{fig:L3} shows that the benefit depends on the shared resource under
contention. When the working set fits in the private L2 cache, bandwidth is not
shared across cores, so desynchronization provides little benefit. The largest
speedup ($\approx70\%$) occurs when the Triad working set fits in the shared L3
cache ($6\times10^{6}$ elements), where desynchronization reduces concurrent L3
bandwidth demand. As the working set grows into main memory, the speedup
decreases monotonically to $\approx5\%$ because memory-bandwidth contention
dominates and the persistent synchronized ccNUMA domain re-emerges. Thus, the
benchmark again requires deliberate noise injection to sustain
desynchronization.

\section{Conclusion and Future Work}
This work studied desynchronization in a regime largely unexplored by prior work:
negligible inter-process communication, with desynchronization instead driven by
the interaction between compute-bound and memory-bound kernels within a single
application. Using the Parallel Optical Flow Solver, we showed that natural
desynchronization improves performance by 25.5\% over a forced-synchronous
baseline, increasing to 46\% for a near-balanced $55\%$ memory-to-compute fraction.
We further showed that removing MPI communication stalls through asynchronous
progress can reduce performance by 80\%, because these stalls unintentionally
limit memory-bandwidth contention.
In contrast, the Pisolver-Triad benchmark does not naturally desynchronize
under main-memory-bound execution and requires periodic ccNUMA-domain-targeted
idle-wave injection to sustain desynchronization, achieving up to 25\% speedup.
However, when the working set fits in shared L3 cache, the benchmark
desynchronizes naturally and achieves substantially higher speedup, showing
that bottleneck characteristics, in addition to multi-kernel structure,
determine desynchronization behavior. Using DisCostiC, we reproduced the
qualitative desynchronization trends of both the benchmark and the full
application, while identifying boundary-rank modeling and secondary idle-wave
generation as key directions for improving simulation fidelity.
These results lead to a simple design rule: maximize performance by keeping the
number of ranks simultaneously executing the memory-bound phase near the
platform's bandwidth-saturation threshold, whether achieved through workload
balance, incidental communication stalls, or deliberate perturbation. A direct
test of this principle is to implement a mechanism that enforces this
condition and compare its performance with the accidental and injected
desynchronization studied here.

Future work will extend this analysis to larger and more complex applications
with multiple interacting kernels, such as climate simulations, and generalize
DisCostiC to support arbitrary multi-kernel workloads while improving trace
fidelity. We will also investigate energy-to-solution, using per-node power
measurements such as LIKWID's RAPL-based energy counters, to determine whether
desynchronization-based runtime improvements translate into energy savings.

\section*{Acknowledgments}
The authors gratefully acknowledge support from the German Research Foundation (DFG), project number 545776403, FOR 5880: Holistic Energy and Performance Modeling for Sustainable Computing (Mod4Comp).
The authors acknowledge the HPC resources provided by the Erlangen National High Performance Computing Center (NHR@FAU) at FAU Erlangen-Nürnberg.
NHR funding is provided by the German Federal Ministry of Education and Research and the state governments participating on the basis of the resolutions of the GWK for the national high-performance computing at universities 
by federal and Bavarian state authorities. 
The NHR@FAU hardware is partially funded by the German Research Foundation (DFG), grant number 440719683.

\bibliographystyle{IEEEtran}
\bibliography{references}

\end{document}